\RequirePackage{fix-cm}
\documentclass[english,5p,preprint,sort&compress]{elsarticle}
\usepackage{lmodern}
\usepackage[T1]{fontenc}
\usepackage[latin9]{inputenc}
\usepackage{array}
\usepackage{booktabs}
\usepackage{fixltx2e}
\usepackage{graphicx}
\usepackage{nomencl}
\usepackage{amsmath}
\providecommand{\printnomenclature}{\printglossary}
\providecommand{\makenomenclature}{\makeglossary}
\makenomenclature

\makeatletter

\providecommand{\tabularnewline}{\\}

\RequirePackage{ifthen}
\renewcommand{\nomgroup}[1]{%
  \ifthenelse{\equal{#1}{Z}}{\item[\textbf{Superscripts}]}{%
  \ifthenelse{\equal{#1}{X}}{\item[\textbf{Subscripts}]}{%
  \ifthenelse{\equal{#1}{G}}{\item[\textbf{Greek symbols}]}{}}}}

\renewcommand\[{\begin{equation}} 
\renewcommand\]{\end{equation}} 

\usepackage{framed} 
\usepackage{multicol} 

\renewcommand*\nompreamble{\begin{multicols}{2}}
\renewcommand*\nompostamble{\end{multicols}}

\usepackage{siunitx}

\@ifundefined{showcaptionsetup}{}{%
 \PassOptionsToPackage{caption=false}{subfig}}
\usepackage{subfig}
\makeatother

\usepackage{babel}
\begin{document}

\begin{frontmatter}{}

\title{Marangoni driven turbulence in high energy surface melting processes}

\tnotetext[t1]{This manuscript has been published in the International Journal of
  Thermal Sciences 104 (June 2016), doi:10.1016/j.ijthermalsci.2016.01.015}

\author[tud,jmbc]{Anton Kidess\corref{cor1}}

\ead{A.Kidess@tudelft.nl}

\author[tud,jmbc]{Sa\v{s}a Kenjere\v{s}}

\author[tud,jmbc]{Bernhard W. Righolt}

\author[tud,jmbc]{Chris R. Kleijn}

\cortext[cor1]{Corresponding author}

\address[tud]{Department of Chemical Engineering, Delft University of Technology,
Julianalaan 136, 2628BL Delft, Netherlands}

\address[jmbc]{JM Burgers Centre for Fluid Mechanics, Mekelweg 2, 2628CD Delft,
Netherlands}
\begin{abstract}
Experimental observations of high-energy surface melting processes,
such as laser welding, have revealed unsteady, often violent, motion
of the free surface of the melt pool. Surprisingly, no similar observations
have been reported in numerical simulation studies of such flows.
Moreover, the published simulation results fail to predict the post-solidification
pool shape without adapting non-physical values for input parameters,
suggesting the neglect of significant physics in the models employed.
The experimentally observed violent flow surface instabilities, scaling
analyses for the occurrence of turbulence in Marangoni driven flows,
and the fact that in simulations transport coefficients generally
have to be increased by an order of magnitude to match experimentally
observed pool shapes, suggest the common assumption of laminar flow
in the pool may not hold, and that the flow is actually turbulent.
Here, we use direct numerical simulations (DNS) to investigate
the role of turbulence in laser melting of a steel alloy with surface
active elements. Our results reveal the presence of two competing
vortices driven by thermocapillary forces towards a local surface
tension maximum. The jet away from this location at the free surface,
separating the two vortices, is found to be unstable and highly oscillatory,
indeed leading to turbulence-like flow in the pool. The resulting
additional heat transport, however, is insufficient to account for
the observed differences in pool shapes between experiment and simulations.\end{abstract}
\begin{keyword}
Marangoni flow\sep Thermocapillary flow\sep Turbulence\sep Direct
numerical simulation \sep Welding
\end{keyword}

\end{frontmatter}{}

\begin{table*}[!t]   \begin{framed}     \printnomenclature   \end{framed} \end{table*}

\section{Introduction}

A long-standing question in the modelling of weld pool hydrodynamics
is the one of the possible occurrence of turbulence and its influence
on heat and momentum transfer. The underlying problem is that no welding
model seems to exhibit true predictive capabilities, not even with
respect to such a simple overall weld pool property as its post-solidification
shape. Rather, all simulations require the adaptation of unphysical
input parameters and/or material properties to truthfully reproduce
experimental results. For instance, \citet{Winkler1997Numerical}
and \citet{Pavlyk2001Numerical} tune the heat input characteristics
as well as the concentration of surface active species to obtain results
matching experiments. More commonly, many authors (e.g. \citep{Anderson2010Origin,De2003Probing,De2004Smart,De2006Improving,Pavlyk2001Numerical,Pitscheneder1996Role,Tan2012Numerical})
resort to the modification (i.e. enhancement) of transport coefficients,
specifically thermal conductivity and viscosity, to match experimental
results. No guideline has been established on how to modify the transport
properties and generally they are tuned on an ad-hoc basis without
any physical reasoning and a priori dependence on weld pool properties.
For example, \citet{Pitscheneder1996Role} enhance the molecular thermal
conductivity and dynamic viscosity by a constant factor 7 to match
experiments, \citet{Anderson2010Origin} increase only the viscosity
by a constant factor 30, \citet{Mishra2008Experimental} increase
only the thermal conductivity by a factor 4, \citet{De2003Probing}
propose an optimization algorithm to determine the best values for
thermal conductivity and viscosity with multiplication factors up
to 17. Even when uncertainties in boundary conditions, e.g. heat transfer
efficiency and energy distribution, are minimal, such as in the conduction-mode
(i.e. with negligible vaporization) laser welding experiments conducted
by \citet{Pitscheneder1996Role}, enhanced transport coefficients
are required to match experimental weld shapes, strongly suggesting
that the published weld pool models lack the inclusion of significant
physics. 

Furthermore, previously published computational studies fail to report
oscillations and non-axisymmetric flow patterns at the liquid surface,
such as have been observed in experiments for conduction-mode laser
and autogeneous gas tungsten arc welds. \citet{Kraus1989Surface}
observes that ``weld pool surface temperature profiles do not reach
quasi-steady-state conditions, but rather vary around some time-averaged
or mean values''. \citet{Zehr1991Thermocapillary} reports that ``high
speed video images of the melt pool seem to reveal substantial oscillations
of the free surface as the laser interacts with the workpiece''.
Finally, \citeauthor{Zhao2009Unsteady} show highly unstable flow
with multiple flow cells using surface particle-image-velocimetry
of a gas-tungsten arc-weld \citep{Zhao2009Unsteady,Zhao2011Measurements}.

A few hypotheses as to how to account for lacking physics, and thus
improve the prediction of weld pool models, have been proposed and
tested by other authors. One identified deficiency is the common comparison
of post-solidification weld pool shapes with numerical simulation
results not including the solidification stage. \citet{Ehlen2003Simulation}
and \citet{Saldi2013Effect} have determined that the weld pool shape
can significantly change during this last stage of a welding process.
Unfortunately, while the inclusion of the solidification stage can
improve the predictions in some situations, it still does not ensure
predictive capabilities \citep{Saldi2013Effect}.

Another possible source of error may be attributed to the often neglected
motion of the liquid-gas interface. Simulations conducted by \citet{Ha2005Study}
based on Pitscheneder's laser welding experiments \citep{Pitscheneder1996Role}
however show a very limited influence of a deformable free surface
on the weld pool shape. The same conclusion has been made by \citet{Zehr1991Thermocapillary}
based on 3D simulations of conduction-mode laser welding.

\citet{Winkler1997Numerical} have proposed the lack of surface chemistry
and surface mass transfer processes in published models, resulting
in a homogeneous distribution of surface active elements such as sulfur
in the pool and at its surface, as potential source of the discrepancy.
The group was able to improve their predictions using a mass transport
model for a surface active element \citep{Winkler2000Effect}, and
even more so when taking into account the effect of multiple surfactants
\citep{Winkler2005Multicomponent}\footnote{It should be noted that \citeauthor{Winkler1997Numerical} use a value
for the standard heat of absorption in disagreement with the commonly
used value \citep{Sahoo1988Surface}, which may have lead to fortuitous
improvement of the results due to a resulting altered surface tension
temperature dependency $d\gamma/dT$.}. However, even though their results using a laminar flow assumption
are promising, they do conclude that there is a need to address the
question of turbulent flow in weld pools. This conclusion is reinforced
by the previously mentioned experimental observations of flow instabilities
which are not seen in the simulations by \citeauthor{Winkler2000Effect}
even when including the effects of surfactant redistribution. 

Although sometimes done without explicit justification (e.g. \citet{He2005Heat,Roy2006Mathematical}),
the hypothesized occurrence of turbulence has been a natural reasoning
for many authors (e.g. \citet{Anderson2010Origin,Choo1994Possible})
to justify increasing transport coefficients, which given turbulent
flow would occur naturally due to turbulent diffusion. A few authors
have attempted to replace the tuning of transport properties by the
use of turbulence models such as RANS \citep{Chakraborty2003Modelling,Chakraborty2004MODELING,Chakraborty2005Influences,Chakraborty2004THREEDIMENSIONAL,Chakraborty2007Modelling,Dong2011GTAW,Goodarzi1998Effect,Hong1998Influence,Hong2002Modelling,Hong2003Vorticity,Jaidi2002kEps,Jaidi2004Threedimensional,Skouras2010Computational,Wang2014Unified}
or LES \citep{Chatterjee2005Largeeddy}. While this leads to improved
agreement with experiments (as does any increase of transport coefficients),
the use of particularly RANS turbulence models developed for aerodynamics
in complexly shaped, Marangoni driven weld pool flows with a free
surface and non-smooth solid-liquid interface, is questionable. In
fact, \citet{Pavlyk2001Numerical} conclude their numerical study
of a gas-tungsten-arc weld with the statement ``that neither an increase
of the transport coefficients by a constant factor nor an application
of the k-$\epsilon$ model improved the correspondence between the
predicted and actual weld pool shapes'', and support further investigation
of the role of turbulence in such flows.

To analyze the possible role of turbulence, \citet{Chakraborty2007Thermal2}
have presented a scaling analysis for high energy surface melting
processes such as the laser welding process of interest here. The
analysis allows the estimation of the flow regime based on three dimensionless
numbers: \emph{(i)} the melt pool depth-to-radius aspect ratio $A=D/L$,
\emph{(ii)} the Prandtl number $Pr$ and \emph{(iii)} a dimensionless
number $N$ inversely proportional to the Marangoni number $Ma$, $
N=\left(\mu\left/\left(\rho\left|{\partial\gamma}/{\partial T}\right|{\eta P}/{(\mu\pi\lambda)}\right)\right.\right)^{1/3}$.

For the Pitscheneder experiment (see table~\ref{tab:Material-properties}
for material properties) at a welding power of \SI{5200}{\watt} and
a sulfur concentration of \SI{150}{ppm}, the values of those dimensionless
numbers are $A\approx1.5$, $Pr=0.178$ and $N\approx0.01$.\nomenclature[aAspect]{$A$}{Aspect ratio}\nomenclature[aDchar]{$D_c$}{Characteristic length scale (pool depth)}\nomenclature[aLchar]{$L_c$}{Characteristic length scale (pool radius)} According to the analysis by Chakraborty and Chakraborty \citep{Chakraborty2007Thermal2}, the onset of turbulence is expected for $2A^{2/3}N^{-2}\geq \mathcal{O}(Re_{crit})$, where $Re_{crit}$ is estimated from experiments to be around $600$ \citep{Chakraborty2007Thermal,Chakraborty2007Thermal2}. Turbulent thermal diffusion is predicted to exceed molecular thermal diffusion when $Pr\geq \mathcal{O}(25N^{2}A^{-2/3})$. Here, $2A^{2/3}N^{-2}\simeq2.6\cdot10^{4}$, and \\$25N^{2}A^{-2/3}\simeq2\cdot10^{-3}$, indicating the flow to be turbulent.

Now that we have established a need to investigate the possibility
of turbulent flow and heat transport in melt pools, we will use simulations
with very high temporal and spatial resolution to investigate the
significance of turbulence without having to resort to questionable
modelling techniques. To date, no such simulation results of welding
have been published, as even with access to supercomputing facilities
the computational cost remains substantial for long welding times.
The stationary conduction-mode laser welding experiments by \citet{Pitscheneder1996Role}
will be used as an attractive test case for the hypothesis of the
occurrence of turbulence, as uncertainties in boundary conditions
are minimized while still exhibiting the need for significantly enhanced
transport coefficients in laminar simulations in order to match the
experimental results. In our simulations, we assume a uniform surfactant distribution in the weld pool, thus focusing on thermal Marangoni effects as a cause for turbulent flow instabilities. Non-uniform surfactant distributions will most likely further contribute to flow instabilities. As such, our present study may be considered as a "best case scenario" for the occurrence of turbulent flow instabilities.

\section{Model formulation}

\subsection{Governing equations}

\begin{figure}
\hfill{}\includegraphics[bb=0bp 0bp 245bp 180bp]{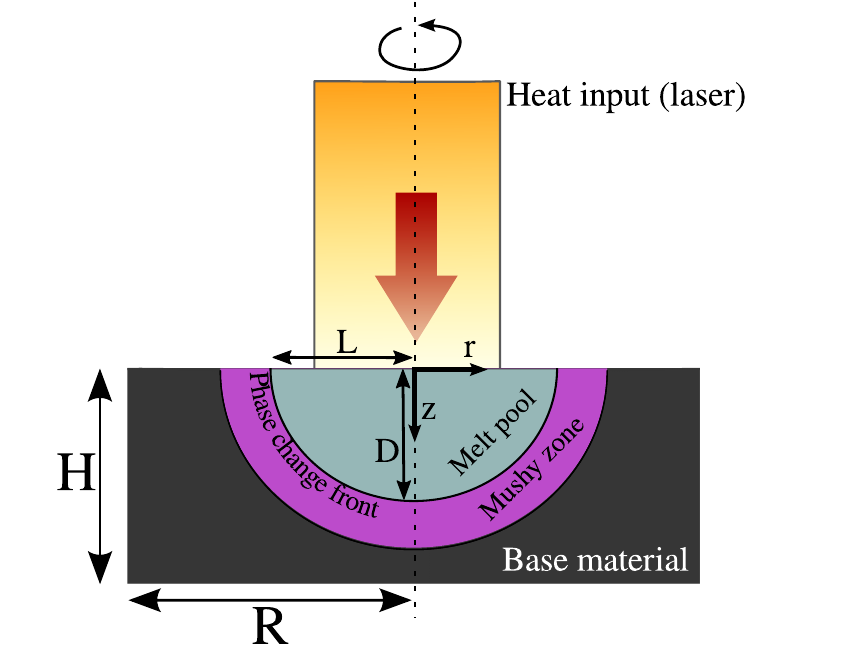}\hfill{}

\protect\caption{Schematic representation of the studied laser welding. \label{fig:Laser-welding-schema}}
\end{figure}

A schematic of a typical weld is shown in figure \ref{fig:Laser-welding-schema},
where a slab of metal is targeted by a high power laser, where the relative speed between the laser and the target is zero. The laser irradiation will be absorbed by the target material, leading to an
increase in temperature and eventually a melting phase change. Heat
will be transferred into the bulk of the welded material by conduction
and thermocapillary driven convection. These phenomena are mathematically
modeled with an energy transport equation with a source term for the
latent heat of the phase change

\begin{equation}
\frac{D}{Dt}(\rho c_{p}T)=\nabla\cdot(\lambda\nabla T)+S_{latent}\label{eq:Energy-equation}
\end{equation}
\nomenclature[aD]{$\frac{D}{Dt}$}{Material derivative}\nomenclature[atime]{$t$}{Time}\nomenclature[grho]{$\rho$}{Density}\nomenclature[acp]{$c_p$}{Heat capacity}\nomenclature[aT]{$T$}{Temperature}\nomenclature[glambda]{$\lambda$}{Thermal conductivity}\nomenclature[aS_latent]{$S_{latent}$}{Latent heat source term}

Due to the non-uniform heating of the top surface, large temperature
gradients will develop. These temperature gradients result in gradients
in surface tension, leading to thermocapillary forces along the non-deformable
liquid-gas interface driving flow in the melt pool. The momentum transport
is described by the Navier-Stokes equations, with a momentum sink
that models the friction in the so-called mushy zone, where the liquid
and solid phase co-exist

\begin{equation}
\frac{D}{Dt}\boldsymbol{u}=-\frac{1}{\rho}\nabla p+\nabla\cdot(\nu\nabla\boldsymbol{u})-\boldsymbol{F}_{damp}\label{eq:Navier-Stokes}
\end{equation}
Here, we have assumed constant density over all phases.

\nomenclature[aU]{$\boldsymbol{u}$}{Fluid velocity}\nomenclature[ap]{$p$}{Pressure}\nomenclature[gmu]{$\mu$}{Dynamic viscosity}\nomenclature[aFdamp]{$\boldsymbol{F}_{damp}$}{Momentum sink term due to solidification}

\subsubsection{Latent heat release}

The effect of melting and solidification on the heat transfer are
taken into account via the source term $S_{latent}$ in equation \ref{eq:Energy-equation}

\begin{equation}
S_{latent}={\displaystyle \rho h_{f}\frac{dg}{dt}}\label{eq:latent-heat-term}
\end{equation}
\nomenclature[aH]{$h_f$}{Latent heat of fusion}\nomenclature[ag]{$g$}{Volume fraction of solid}

with $g$ the volume fraction of solid material, which is assumed
to vary linearly over the melting temperature range between solidus
and liquidus 

\begin{equation}
g=\frac{T_{l}-T}{T_{l}-T_{s}},\,T_{s}<T<T_{l}\label{eq:latent-heat-solid-fraction-fs}
\end{equation}

\nomenclature[aTsl]{$T_s$, $T_l$}{Solidus and liquidus temperature}

\subsubsection{Coupling of momentum and heat transport}

Through the inclusion of the momentum sink term, the momentum equation
\ref{eq:Navier-Stokes} is valid for the entire domain including both
liquid and solid regions. The (semi-)solid regions are modeled as
a porous medium, introducing a momentum sink following the isotropic
Blake-Kozeny model~\citep{Singh2001Modelling}

\[
\boldsymbol{F}_{damp}=\frac{\mu}{K\rho}\boldsymbol{u}
\]

\[
K=K_{0}\frac{g^{2}}{(1-g)^{3}+\varepsilon}
\]
with $\mu/K_{0}=\SI{e6}{\newton\second\per\meter\tothe{4}}$ and $\varepsilon=10^{-3}$. A similar approach has successfully been applied in DNS by \citet{Breugem2006Influence}.

\subsection{Boundary conditions}

For 2D simulations we assume the melt pool to be axisymmetrical and
make use of this by only simulating a wedge of the domain. Circumferential
gradients are zero on the wedge faces. The conditions on the remaining
boundaries (which are the same in 2D and 3D) are outlined in the following.

\subsubsection{Heat input}

At the top surface, the laser irradiation is modeled by a top-hat
distributed heat flux. Because the heat loss due to radiation and
convection is only a small fraction of the laser irradiation, we apply
adiabatic boundary conditions everywhere except the irradiated area,
where we apply a top-hat distribution as

\begin{equation}
\lambda\nabla_{n}T\Big|_{z=0}=\frac{\eta P}{\pi r_{q}^{2}},\,r\leq r_{q}\label{eq:bc_laser_heat}
\end{equation}

\nomenclature[geta]{$\eta$}{Laser absorptivity}\nomenclature[aP]{$P$}{Laser power}\nomenclature[arq]{$r_q$}{Laser beam radius}

Here we follow \citet{Pitscheneder1996Role} with $\eta=0.13$, $P=\SI{5200}{\watt}$
and $r_q=\SI{1.4}{\milli\meter}$.

\subsubsection{Momentum}

At the liquid-gas interface, we introduce a shear stress in the liquid
due to surface tension gradients along the interface (Marangoni force):

\begin{equation}
\mu\nabla_{n}u_{t}\Big|_{z=0}=\frac{d\gamma}{dT}\nabla_{t}T\label{eq:Marangoni-BC}
\end{equation}

\begin{figure}
\includegraphics{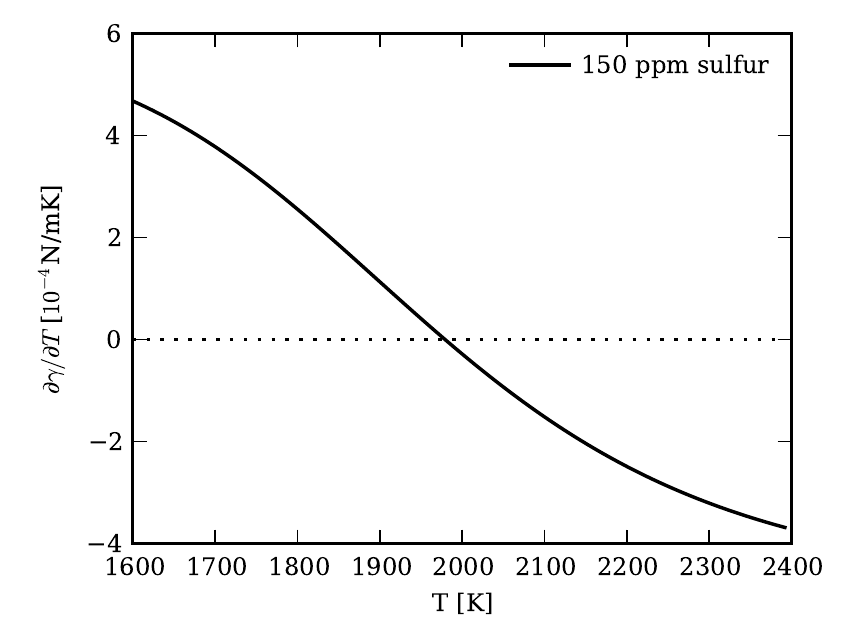}

\protect\caption{Surface tension temperature coefficient\label{fig:dsdT-150ppm}}
\end{figure}

The variation of surface tension with temperature is computed using
the thermochemical model of \citet{Sahoo1988Surface}. The relevant
curve for a sulfur concentration of \SI{150}{ppm} is plotted in figure~\ref{fig:dsdT-150ppm}. Experimentally, such a behaviour with a sign change at a critical temperature has been shown to occur in steels \citep{Mills1990Factors,Ozawa2014InfluenceS} and pure iron \citep{Hibiya2013Effect}, as well as other metals such as silver \citep{Hibiya2013Effect} and nickel \citep{Ozawa2014Influence}.

Based on the small Capillary number $Ca=\mathcal{O}(\num{e-2})$ for the studied weld pool, indicating that surface tension will effectively counter-act deformations due to fluid flow, we assume the free surface to be non-deformable. This assumption is in line with the observations by \citet{Ha2005Study}, who investigated the influence of free surface deformations for the \citet{Pitscheneder1996Role} case and concluded it is not important. The non-deformable surface assumption, however, may not hold for other welding  processes and conditions at higher Capillary numbers, as experimental results show \citep{kou2011oscillatory,Zhao2009Unsteady,Zhao2011Measurements}.

At all other surfaces, we set the velocity to zero. \nomenclature[ggamma]{$\gamma$}{Surface tension}\nomenclature[xsubscriptt]{$t$}{Tangential direction}\nomenclature[xsubscriptn]{$n$}{Normal direction}

\section{Numerical procedure}

Our solver is built on top of the open source finite volume framework
OpenFOAM (version 2.1.x) \citep{Weller1998Tensorial}. 

We use a 2nd order backward differencing time marching scheme, and
a 2nd order TVD scheme (limitedLinear \citep{Berberovic2010Investigation})
for the divergence terms. At every time step, the non-linearity associated
with the pressure-velocity-coupling is handled by the iterative PISO
algorithm \citep{Issa1986Solution}. Once a divergence free velocity
field has been computed at a given time step, the energy transport
equation (\ref{eq:Energy-equation}) is solved. If a phase change
occurs, the temperature equation will be non-linear. The non-linearity
due to latent heat is dealt with using an implicit source term linearization
technique \citep{Voller1991GENERAL}.\footnote{The solver and input files will be made available through the journal's
supplementary material.} 

\begin{table}
\protect\caption{Material properties of the S705 alloy \citep{Pitscheneder1996Role}\label{tab:Material-properties}}

{\small{}}%
\begin{tabular*}{8.5cm}{@{\extracolsep{\fill}}>{\raggedright}b{4.2cm}r>{\raggedright}p{1.4cm}}
\toprule 
{\small{}Property} & {\small{}Value} & {\small{}Unit}\tabularnewline
\midrule
{\small{}Solidus temperature $T_{s}$} & {\small{}$1610$} & {\small{}\si{\kelvin}}\tabularnewline
{\small{}Liquidus temperature $T_{l}$} & {\small{}$1620$} & {\small{}\si{\kelvin}}\tabularnewline
{\small{}Specific heat capacity $c_{p}$} & {\small{}$670$} & {\small{}\si{\joule\per\kilo\gram\per\kelvin}}\tabularnewline
{\small{}Density $\rho$} & {\small{}$8100$} & {\small{}\si{\kilo\gram\per\cubic\meter}}\tabularnewline
{\small{}Thermal conductivity $\lambda$} & {\small{}$22.9$} & {\small{}\si{\watt\per\meter\per\kelvin}}\tabularnewline
{\small{}Latent heat of fusion $h_{f}$} & {\small{}$2.508\cdot10^{5}$} & {\small{}\si{\joule\per\kilo\gram}}\tabularnewline
{\small{}Viscosity $\mu$} & {\small{}$6\cdot10^{-3}$} & {\small{}\si{\pascal\second}}\tabularnewline
{\small{}Surface tension temperature coefficient $\partial\gamma/\partial T|_0$} & {\small{}$-5.0\cdot10^{-4}$} & {\small{}\si{\newton\per\meter\per\kelvin}}\tabularnewline
{\small{}Entropy factor} & {\small{}$3.18\cdot10^{-3}$} & {\small{}$-$}\tabularnewline
{\small{}Standard heat of adsorption} & {\small{}$-1.66\cdot10^{8}$} & {\small{}\si{\joule\per\kilo\mol}}\tabularnewline
{\small{}Surface excess at saturation} & {\small{}$1.3\cdot10^{-8}$} & {\small{}\si{\kilo\mol\per\square\meter}}\tabularnewline
\bottomrule
\end{tabular*}{\small \par}
\end{table}

To properly resolve the turbulent structures in space and time using
direct numerical simulations, we estimate the length and time scales
of the smallest turbulent eddies (Kolmogorov scales), which depend
on a characteristic velocity and a characteristic length scale. Looking
at the experimental and numerical results reported by \citet{Pitscheneder1996Role},
we estimate a characteristic velocity $U_{c}\approx\SI{0.2}{\meter\per\second}$,
and a characteristic length scale of $2L_{c}\approx\SI{4e-3}{\meter}$.
Now with the turbulent kinetic energy dissipation rate $\epsilon\approx U_{c}^{3}/D$,
the Kolmogorov length scale is estimated by\nomenclature[gnu]{$\nu$}{Kinematic viscosity}\nomenclature[gepsilon]{$\epsilon$}{Turbulent kinetic energy dissipation rate}\nomenclature[aLKolmogorov]{$L_K$}{Kolmogorov length scale}\nomenclature[aU]{$U_{c}$}{Characteristic velocity}

\begin{equation}
L_{K}=\left(\frac{\nu^{3}}{\epsilon}\right)^{1/4}=\left(\frac{D\nu^{3}}{U_{c}^{3}}\right)^{1/4}\approx\SI{2e-5}{\meter}\label{eq:Kolmogorov-length-scale}
\end{equation}
\nomenclature[atKolmogorov]{$t_K$}{Kolmogorov time scale}The Kolmogorov
time scale is given by

\[
t_{K}=\left(\frac{\nu}{\epsilon}\right)^{1/2}\approx\SI{6e-4}{\second}
\]

\begin{figure}
\includegraphics{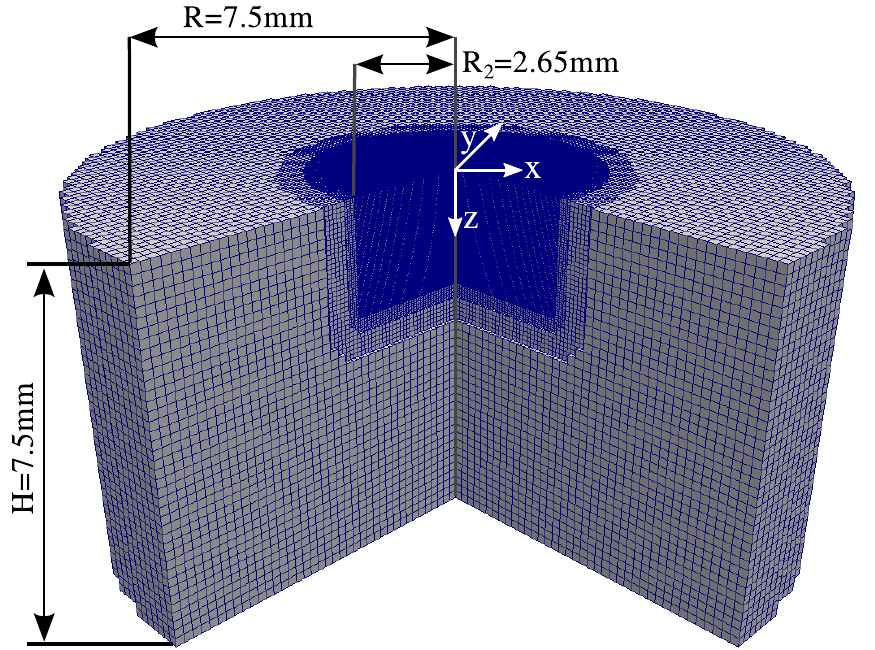}

\protect\caption{3D mesh, where one quarter of the domain has been clipped for visualization.
The coarse outer mesh with a grid cell size of \SI{188}{\micro\meter}
is refined in three steps to the finest inner mesh with a grid cell
size of \SI{23}{\micro\meter}. The latter is too fine to be resolved
in this figure.\label{fig:3D-mesh}}
\end{figure}

The solution domain is a cylinder of radius $R=\SI{7.5}{\milli\meter}$
and height $H=\SI{7.5}{\milli\meter}$, discretized with a mesh of
4.8 million cubic control volumes. The area where we expect fluid
flow consists of small cubes with a a cell spacing of \SI{23}{\micro\meter},
whereas we use larger cells of \SI{188}{\micro\meter} away from
the liquid region. The mesh is shown in figure~\ref{fig:3D-mesh}.
The time step is dynamically set obeying a maximum Courant number
of $Co=U\Delta t/\Delta x<0.33$, resulting in a typical time step
of less than \SI{1e-5}{\second}.

\begin{figure}
\includegraphics[width=3.5in]{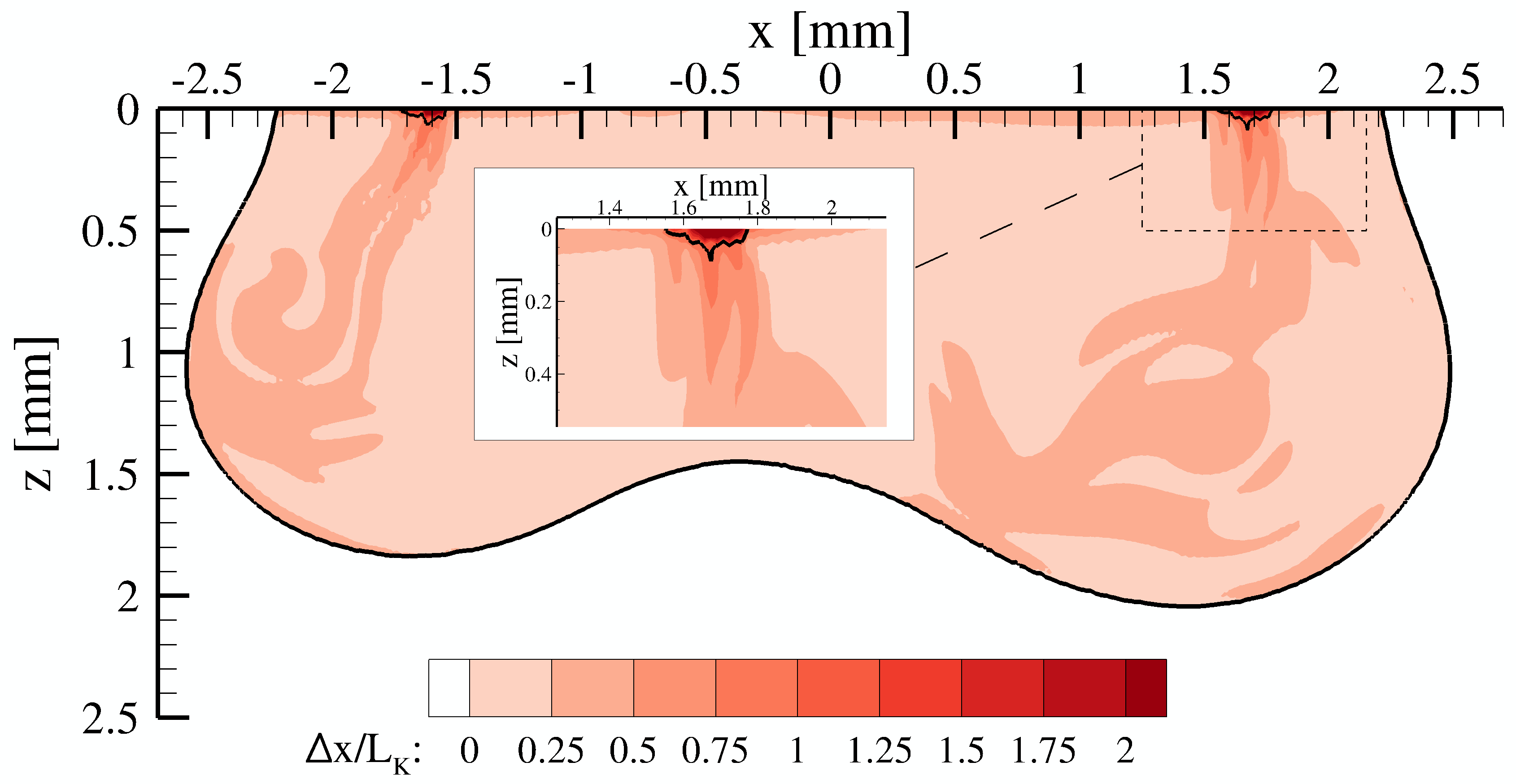}

\protect\caption{Grid size to turbulent length scale ratio $\Delta x/L_{K}$ in the
in the y=\SI{0}{\milli\meter} plane, with the dissipation $\epsilon$
averaged over a time of 0.5 seconds. A length scale ratio smaller
than 1 means turbulence is perfectly resolved, which is the case everywhere
except a small region near the stagnation point at the surface (separated
by a black line, with a maximum value of 4). This and subsequent figures show (quasi) instantaneous cross sections of the strongly unstable and non-axisymmetric weld pool. Such cross sectional snapshots are therefore presented in the x-z plane at y=0 of a Cartesian coordinate system, rather than the r-z coordinate system used in figures 1-3.\label{fig:Length-scale-ratio}}
\end{figure}

To further demonstrate the sufficient resolution of our mesh for proper
direct numerical simulation (DNS) of the liquid, anticipating the simulation results presented in the next section, we determine the distribution of the
turbulence dissipation rate in the simulated flow as $\epsilon=\nu\overline{\nabla \boldsymbol{u}':\nabla \boldsymbol{u}'}$,
with the computed velocity fluctuations $\boldsymbol{u}'=\boldsymbol{u}-\overline{\boldsymbol{u}}$\nomenclature[aUp]{$u'$}{Velocity fluctuation}\nomenclature[aUm]{$\overline{\boldsymbol{u}}$}{Mean velocity}.
The ratio of the mesh spacing $\Delta x$ and the smallest turbulence
length scales $L_{K}$, based on the simulated velocity and dissipation
averaged over a time of \SI{0.5}{\second}, is plotted in figure~\ref{fig:Length-scale-ratio}
for a slice through the pool, showing excellent resolution of even
the smallest scales in our simulations. Only a very small region near
the stagnation point at the surface, consisting of few mesh cells,
is under-resolved by a factor up to 4.

\section{Results and Discussion}

\subsection{Verification with enhanced transport coefficients}

\begin{figure}
\includegraphics{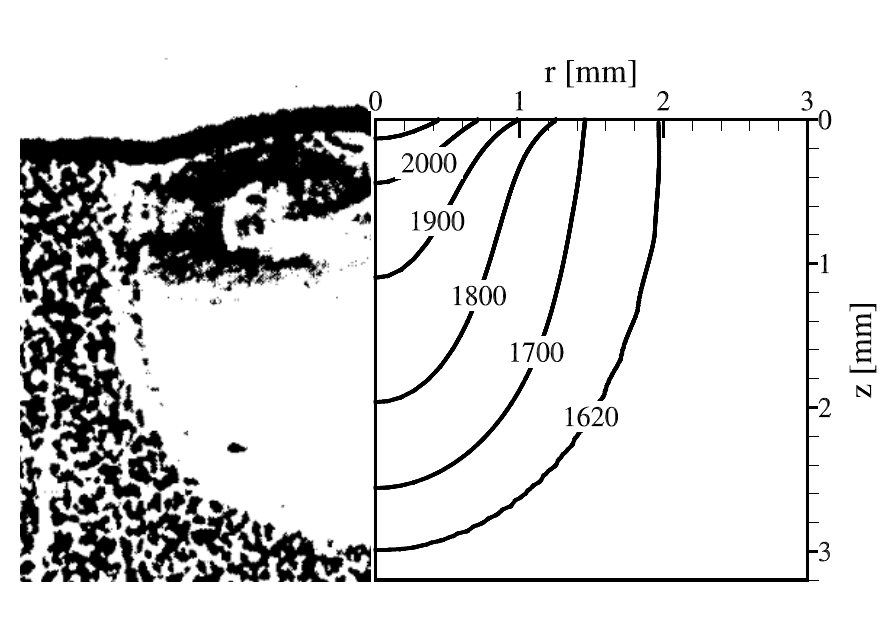}

\protect\caption{The right hand side of the figure shows the current faithful representation
of the experimental result (left half of the figure, reproduced from
\citep{Pitscheneder1996Role} with permission of the publisher, sized to match scale on the right) after
\SI{5}{\second} of welding, using an enhancement factor of 7 for
both the viscosity and the thermal conductivity. Temperature isolines
in Kelvin (simulation result, right half).\label{fig:PitschF7-exp}}

\end{figure}

In order to obtain good agreement between their numerically simulated
melt pool shapes and experimentally observed post-solidification weld
shapes, \citet{Pitscheneder1996Role} artificially enhance the viscosity
and thermal conductivity of the welded steel by a non-elucidated constant
factor 7. We can reproduce their result using this constant enhancement
factor, when, as done by \citeauthor{Pitscheneder1996Role}, we use
a coarse 2D-axisymmetric grid\footnote{Strongly refined towards the free surface and pool centre, with the
smallest $\Delta{}r=\SI{90}{\micro\meter}$ and $\Delta{}z=\SI{12}{\micro\meter}$}, relatively large time steps, and a diffusive upwind discretization
scheme (see figure~\ref{fig:PitschF7-exp}). Without artificially
increasing the transport coefficients, i.e. when using physically
realistic values for the viscosity and the thermal conductivity, the
flow within the melt pool differs significantly and so does the obtained
final weld pool shape, as we will show in the following section.

\subsection{Direct numerical simulations without enhancement of transport properties}

\begin{figure}
\hfill{}\includegraphics[clip,width=3.55in]{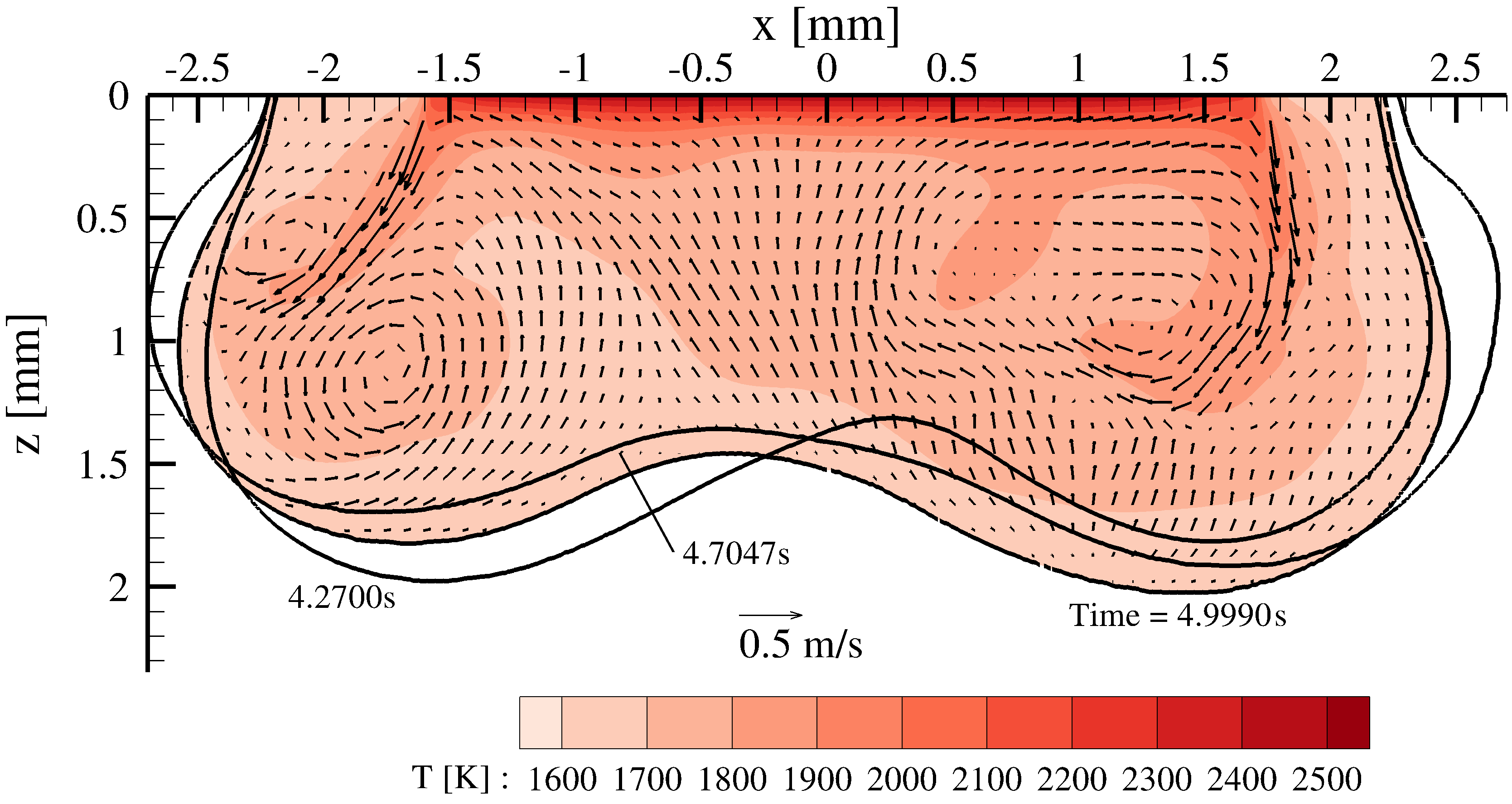}\hfill{}

\protect\caption{Melt pool shape, temperatures (in Kelvin) and velocity vectors in
the y=\SI{0}{\milli\meter} plane at the end of heating $t=\SI{5.0}{\second}$.
Note that the velocity vectors have been interpolated to a coarse
grid in post-processing for clarity. The pool shape at two other time
instances is overlayed, showing the pool boundary oscillation. \label{fig:3D-weld-pool-shape}
}
\end{figure}

The melt pool shape after \SI{5.00}{\second} of heat input, obtained from three-dimensional direct numerical simulations with realistic (non-enhanced) transport properties, is shown in figure~\ref{fig:3D-weld-pool-shape}.
Also shown are melt pool shape snapshots after \SI{4.27}{\second} and \SI{4.70}{\second} of heat input. Compared to the results obtained with enhanced transport coefficients and a diffusive numerical scheme on a coarse 2D mesh, as shown in figure~\ref{fig:PitschF7-exp}, it is now clearly visible that \emph{(i)} the flow has not remained symmetric, leading to an asymmetric melt pool shape at this time instance; \emph{(ii)} the melt pool is a bit wider and much less deep, leading
to a pool depth-to-radius aspect ratio which is now smaller than 1; \emph{(iii)} The melt pool shape is now strongly time dependent and oscillating.

These observations may be understood as follows: Due to the low (i.e. non-enhanced) molecular thermal conductivity, higher temperatures are now sustained at the melt pool surface, which lead to a large region subject to a negative surface tension gradient, in place of
the previously dominant positive surface tension gradient (figure~\ref{fig:dsdT-150ppm}).
The latter caused a flow directed inward along the pool surface, and towards the pool bottom along its axis, resulting in a deep, hemispherical pool shape as shown in figure~\ref{fig:PitschF7-exp}. The sign change in the surface tension gradient now leads to a surface flow directed
radially outward from the pool centre, rather than the inward directed flow in the 2D simulation with enhanced transport coefficients. This results in a wide, shallow pool, rather than a deep, narrow pool. At the edge of the melt pool surface, where temperatures are lower,
the surface tension gradient is still positive. As a result, the radially outward surface flow from the pool centre impinges onto a second surface flow directed radially inward from the edge of the pool towards the pool centre. At the point where the two opposing flows meet, at a
radial distance of roughly \SI{1.5}{\milli\meter}, a circular, downward jet from the pool surface towards the base of the pool is formed. The downward jet is unstable, as both its origin and its angle oscillate in time. We use the term "instability" as it stresses that the initially laminar flow has transitioned into a chaotic state, and not a mere regular laminar unsteadiness. The general flow topology has been anticipated by \citet{Mills1998Marangoni} and \citet{Keene1982Effects}. \citeauthor{Mills1998Marangoni} also comment on the possibility of thermocapillary instabilities arising due to temperature gradients normal to the free surface, based on a theory formulated by \citet{Nemchinsky1997Role} assuming constant $\partial\gamma/\partial T$. Here however, in contrast to the case of \citeauthor{Nemchinsky1997Role}, the downward jet is clearly the dominating source of turbulent motion, as opposed to capillary waves at the free surface. The oscillating downward jet due to the sign change in surface tension also sets the present case apart from previous investigations of thermocapillary instabilities with constant, negative $\partial\gamma/\partial T$  \citep{Kuhlmann2010Flow,Karcher2000Turbulent,Boeck2003LowPrandtlNumber,Dikshit2009Convection}.

\begin{figure}
\hfill{}\includegraphics[width=3.55in]{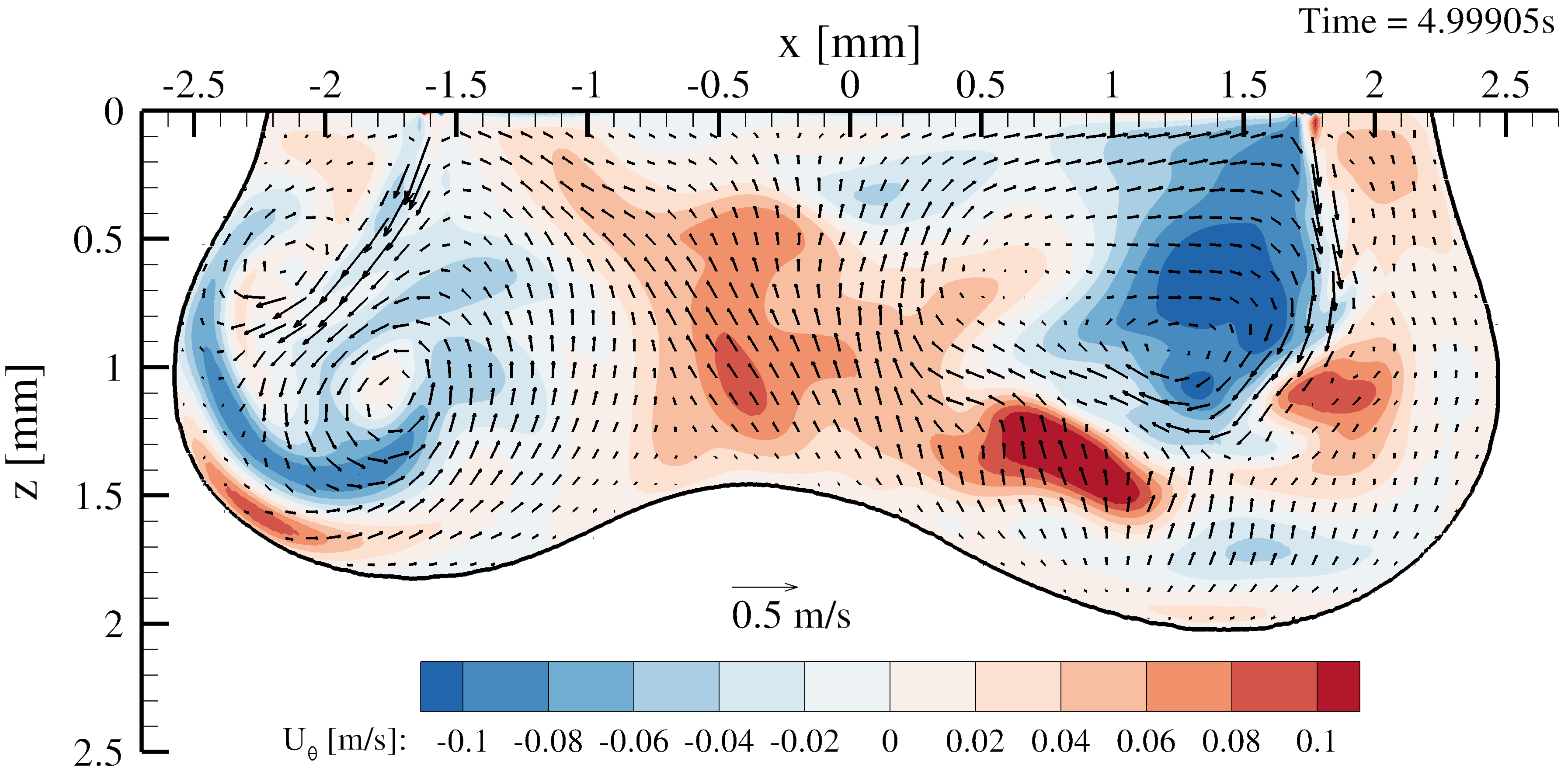}\hfill{}

\protect\caption{In-plane (x,z) velocities in the y=\SI{0}{\milli\meter} plane at
$t=\SI{5.0}{\second}$ indicated by vectors, and out-of-plane (azimuthal)
velocities indicated by colour contours.\label{fig:3d-out-of-plane-flow}}
\end{figure}

\begin{figure*}
\hfill{}\includegraphics[clip,width=7in]{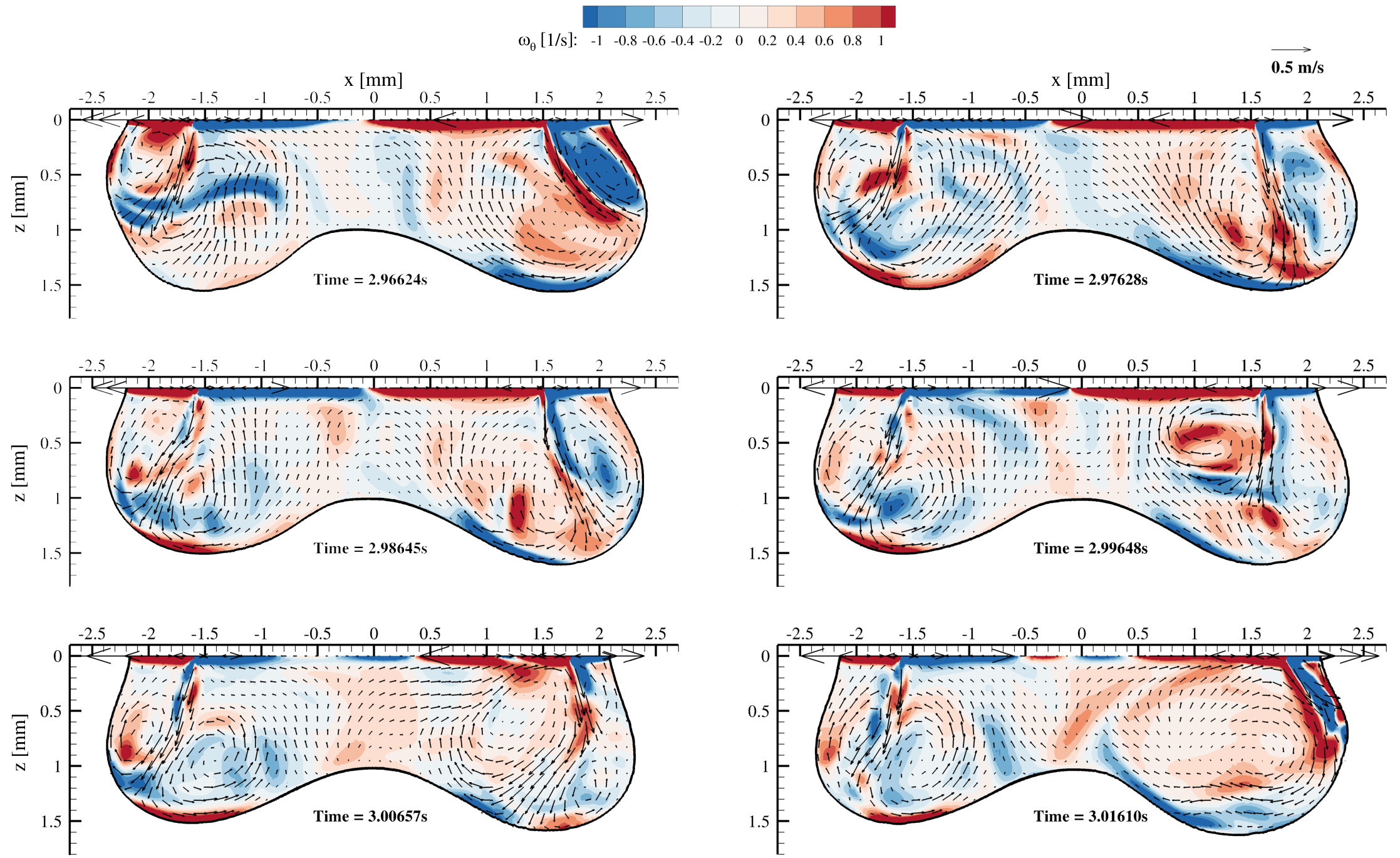}\hfill{}

\protect\caption{In-plane (x,z) velocity vectors in the y=\SI{0}{\milli\meter} plane
at a few time instances around $t=\SI{3.0}{\second}$ (approximately
\SI{0.01}{\second} apart), and out-of-plane vorticity contours.\label{fig:3d-dns-vort}}
\end{figure*}

The oscillating, hot, downwardly directed jet transports so
much heat away from the surface that the melt pool boundary is continuously melting and re-solidifying, depending on where the jet is facing at a given time instance. This causes the oscillation of the pool boundary, as visible from the overlayed pool shapes at two additional time instances in figure~\ref{fig:3D-weld-pool-shape}. The flow is also strongly three dimensional, with significant, unsteady flow present in the azimuthal direction (figure~\ref{fig:3d-out-of-plane-flow}). The
flow pattern and vorticity $\omega$\nomenclature[gomega]{$\omega$}{Vorticity} at various time instances around $t=\SI{3.0}{\second}$, roughly \SI{0.01}{\second} apart, is shown in figure~\ref{fig:3d-dns-vort}. During these time instances the right jet oscillates back and forth, whereas the left jet is relatively stable. This is of course not true for all time instances, highlighting the chaotic nature of the independent motion of the two jets.

\begin{figure*}
\hfill{}\includegraphics[bb=0bp 0bp 2334bp 861bp,clip,width=9in]{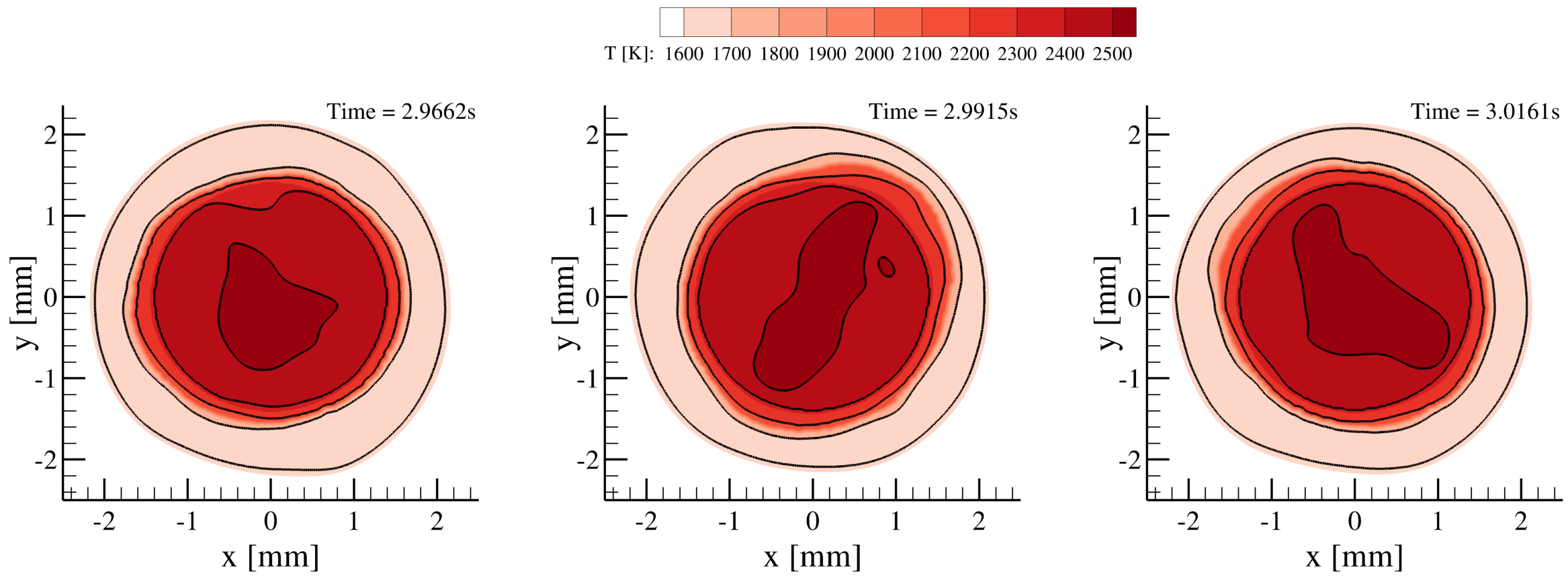}\hfill{}

\protect\caption{Pool surface (z=\SI{0}{\milli\meter}) temperatures in Kelvin at three
time instances, top view. Isolines are drawn at 1620K, 1700K, 2200K,
2400K and 2500K (from outside to inside).\label{fig:3D-weld-pool-top-view}}
\end{figure*}

The melt pool flow instabilities are also visible in the temperatures
at the pool surface, shown in figure~\ref{fig:3D-weld-pool-top-view}.
The oscillations are most apparent in the centre of the pool where
the oscillation frequency is high, but also at the rim of the pool
with a lower frequency due to the dampening effect of melting and
re-solidification. 

\begin{figure*}
\hfill{}\includegraphics[bb=0bp 0bp 2334bp 861bp,clip,width=9in]{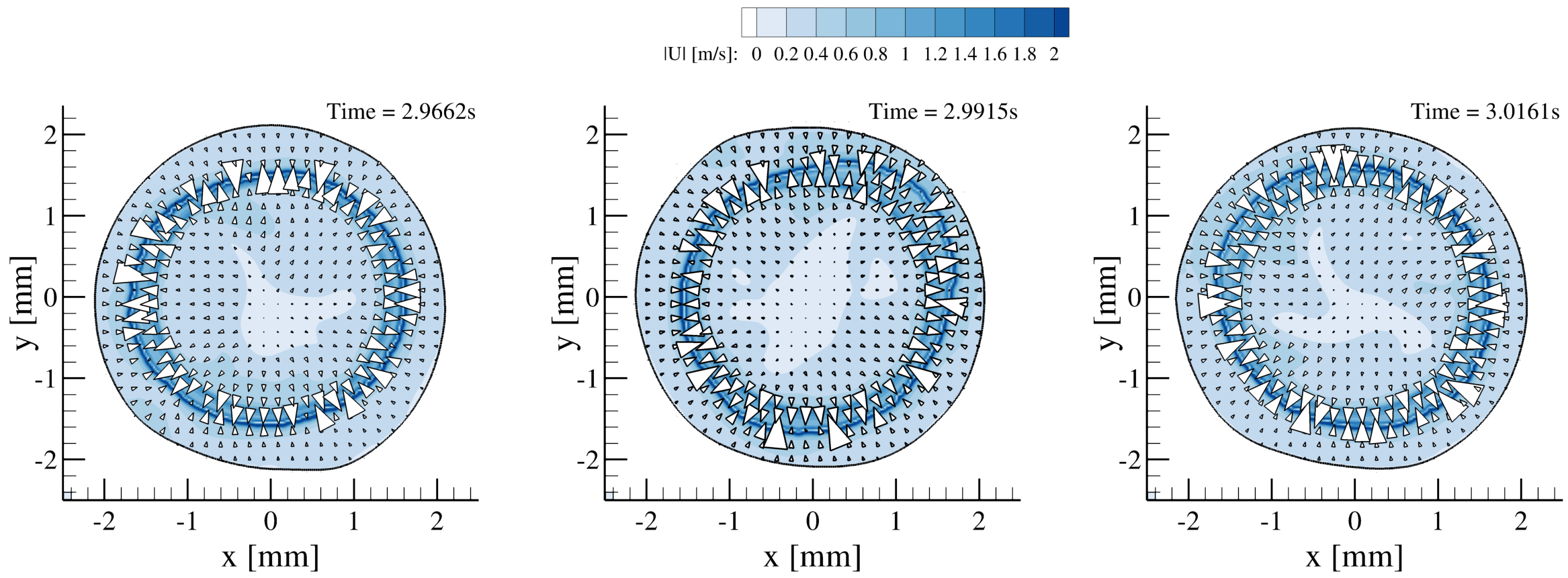}\hfill{}

\protect\caption{Pool surface (z=\SI{0}{\milli\meter}) flow at three time instances,
top view. The largest vectors in the vicinity of the stagnation point
have been blanked for clarity.\label{fig:3D-weld-pool-top-view-velmag}}
\end{figure*}

At the stagnation line, where the radially outward surface flow from
the pool centre impinges on the radially inward surface flow from
the pool edge and where there is a sign change in the surface tension
coefficient, very high thermal gradients of O(\SI{3000}{\kelvin\per\milli\meter})
occur. Since the thermocapillary force is proportional to these thermal
gradients, this is also where we encounter the highest flow velocities
(figure~\ref{fig:3D-weld-pool-top-view-velmag}), locally as high
as \SI{2}{\meter\per\second}.

\subsection{Analysis of turbulent flow properties}

We now address a more quantitative analysis of the turbulent nature
of the melt pool flow, and the importance of turbulent heat transfer. 

\begin{figure}
\includegraphics[width=3.55in]{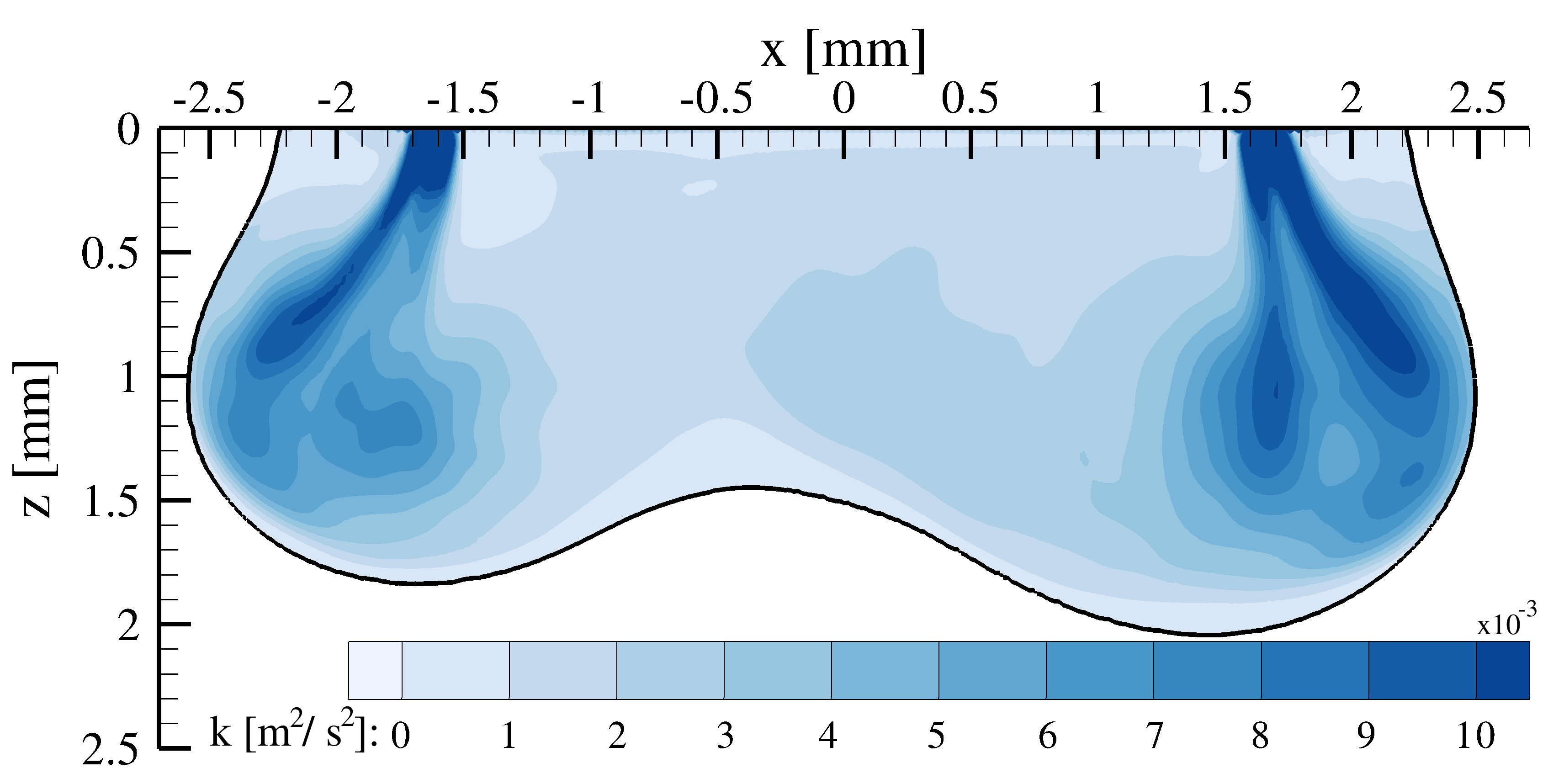}

\protect\caption{Turbulent kinetic energy $k$ in the y=\SI{0}{\milli\meter} plane,
averaged over \SI{0.5}{\second}.\label{fig:Turbulent-kinetic-energy}}

\end{figure}

\begin{figure}
\includegraphics[width=3.55in]{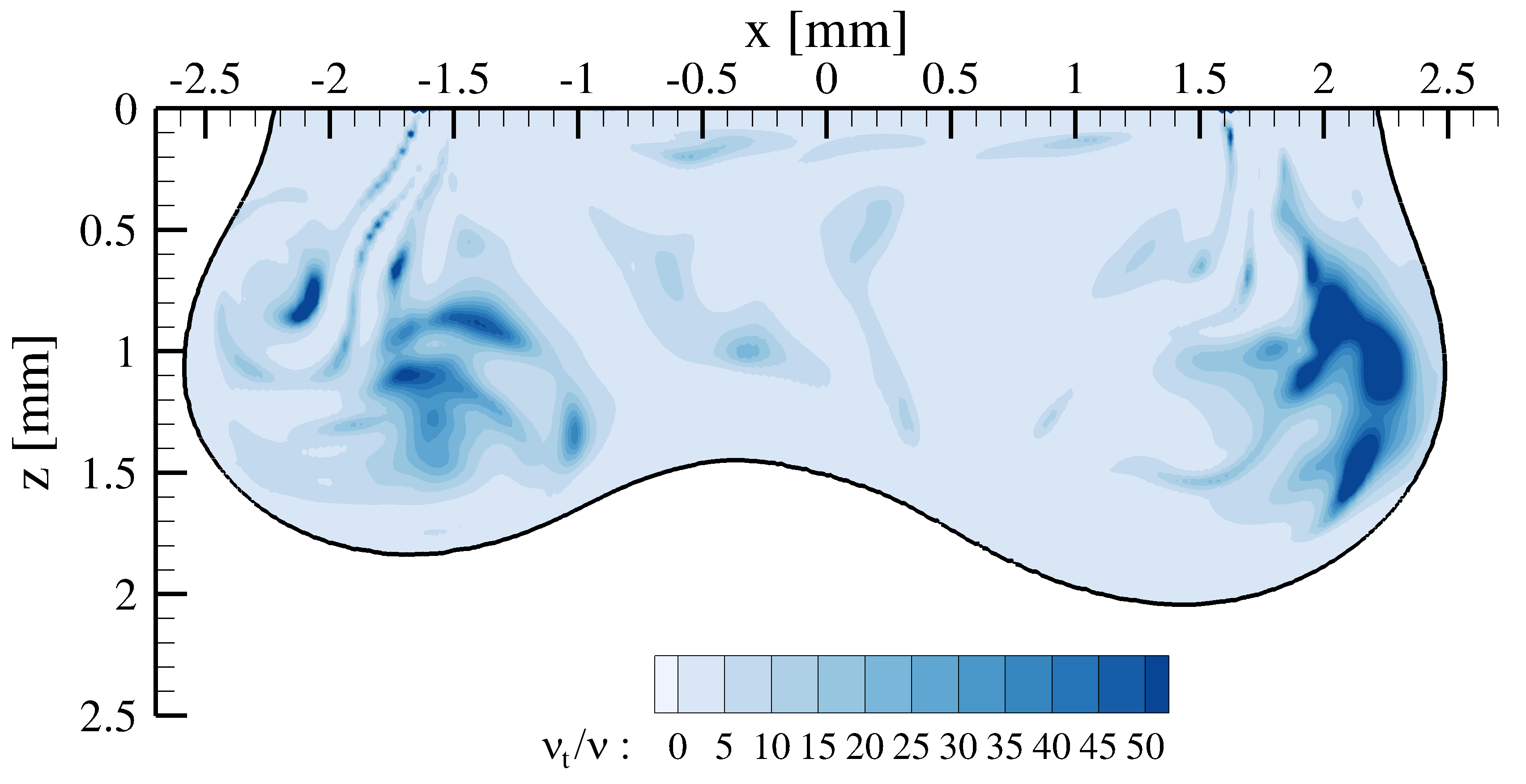}

\protect\caption{Ratio of turbulent diffusivity over molecular diffusivity $\nu_{t}/\nu$
in the y=\SI{0}{\milli\meter} plane, based on turbulent kinetic energy
and turbulence dissipation averaged over \SI{0.5}{\second}.\label{fig:nu_turb_over_nu}}
\end{figure}

Using the computed instantaneous velocity fluctuations $\boldsymbol{u}'=\boldsymbol{u}-\bar{\boldsymbol{u}}$,
we can determine the turbulent kinetic energy $k=\overline{\boldsymbol{u}'\cdot \boldsymbol{u}'}/2$\nomenclature[ak]{$k$}{Turbulent kinetic energy},
and the turbulent viscosity as $\nu_{t}=0.09k^{2}/\epsilon$, with
the turbulent kinetic energy dissipation rate $\epsilon=\nu\overline{\nabla \boldsymbol{u}':\nabla \boldsymbol{u}'}$.
Here, all averages have been computed over the time interval between
4.5 and 5\si{\second}. The results are shown in figures \ref{fig:Turbulent-kinetic-energy}
and \ref{fig:nu_turb_over_nu}, respectively. The turbulent kinetic
energy takes its highest values near the extreme positions of the
jet and near the stagnation point at the free surface. The turbulent
viscosity assumes its largest values of roughly 50 times the molecular
value in an area around the end point of the jet. The space averaged
value of the turbulent viscosity is approximately 7.4 times the molecular
value. This, coincidentally, is close to the factor 7 enhancement
for the transport properties used by \citet{Pitscheneder1996Role}
to match their experimental results. However, the uniform enhancement
used by \citeauthor{Pitscheneder1996Role} leads to distinctly different
melt pool shapes than the turbulent enhancement following from our
DNS simulations. In the first, a hemispherical melt pool shape is
obtained which is deepest at the centre, whereas the maximum turbulent
enhancement occurs in the oscillating jet regions and causes the pool
to be wider and deeper at the edges.

\begin{figure}
\subfloat[Temperature at monitoring point.\label{fig:3D-temperature-probe}]{\hfill{}\protect\includegraphics{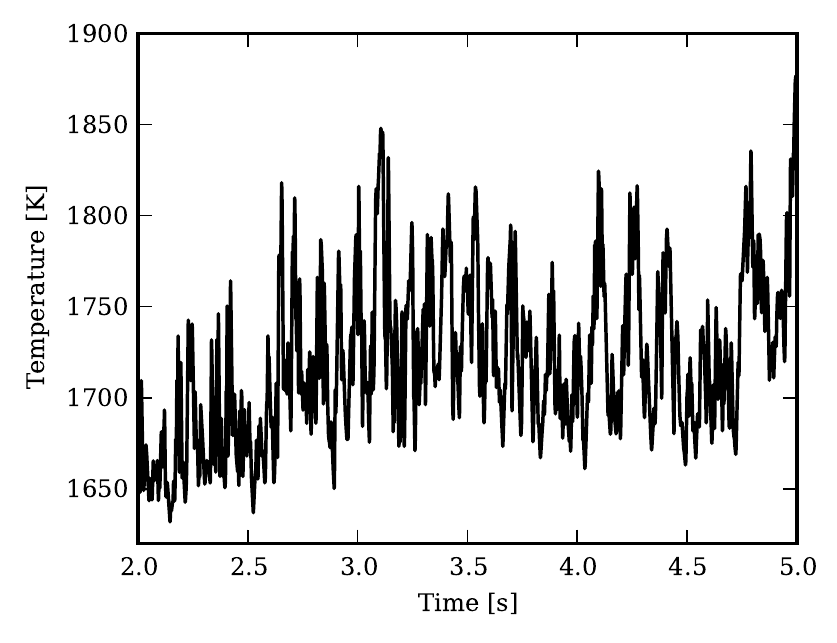}\hfill{}

}

\subfloat[Total velocity magnitude at monitoring point.\label{fig:3D-velmag-probe}]{\hfill{}\protect\includegraphics{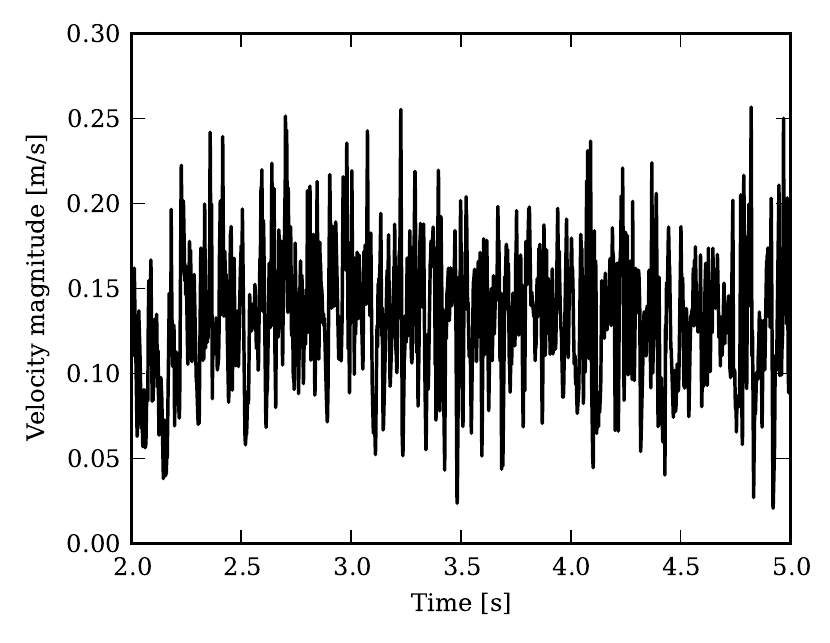}\hfill{}

}

\protect\caption{Temperature and velocity magnitude at monitoring location x=\SI{1}{\milli\meter},
y=\SI{0}{\milli\meter}, z=\SI{1}{\milli\meter}\label{fig:3D-Temperature-monitoring-points}}
\end{figure}

To further quantitatively investigate the oscillating flow, we track
a monitoring point at a distance of x=\SI{1}{\milli\meter} and a
depth of z=\SI{1}{\milli\meter} from the centre of the pool surface.
The temperature history at this monitoring point is shown in figure~\ref{fig:3D-temperature-probe}.
After reaching a quasi steady state, it shows an irregular oscillation
with an amplitude of about \SI{200}{\kelvin} around a mean temperature
of \SI{1720}{\kelvin}. The velocity magnitude at the monitoring point
(figure~\ref{fig:3D-velmag-probe}) oscillates violently with an
amplitudes of roughly 50\% of its mean value. 

\begin{figure}
\subfloat[DFT of temperature at monitoring point\label{fig:3D-temperature-probe-fft}]{\hfill{}\protect\includegraphics{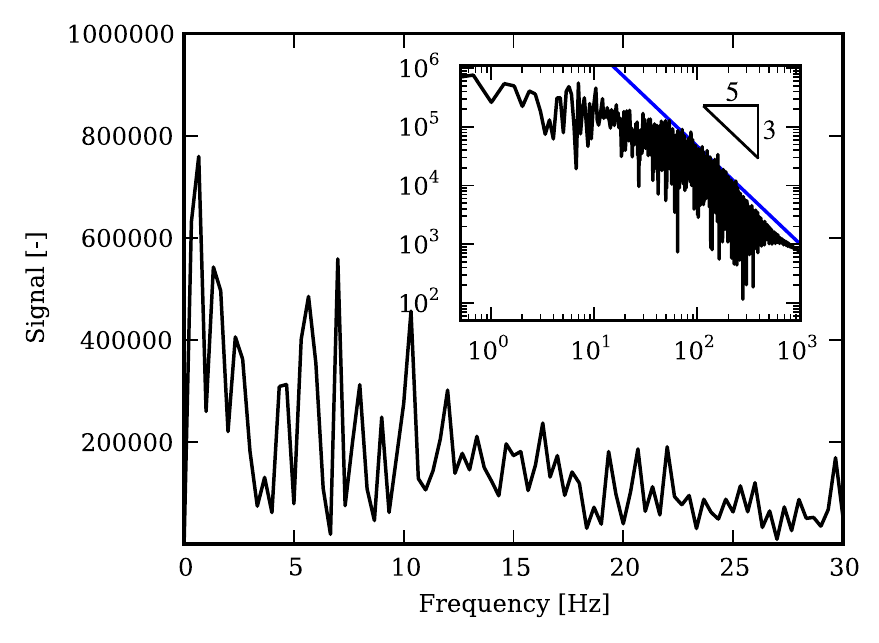}\hfill{}

}

\subfloat[DFT of velocity magnitude at monitoring point\label{fig:3D-velmag-probe-fft}]{\hfill{}\protect\includegraphics{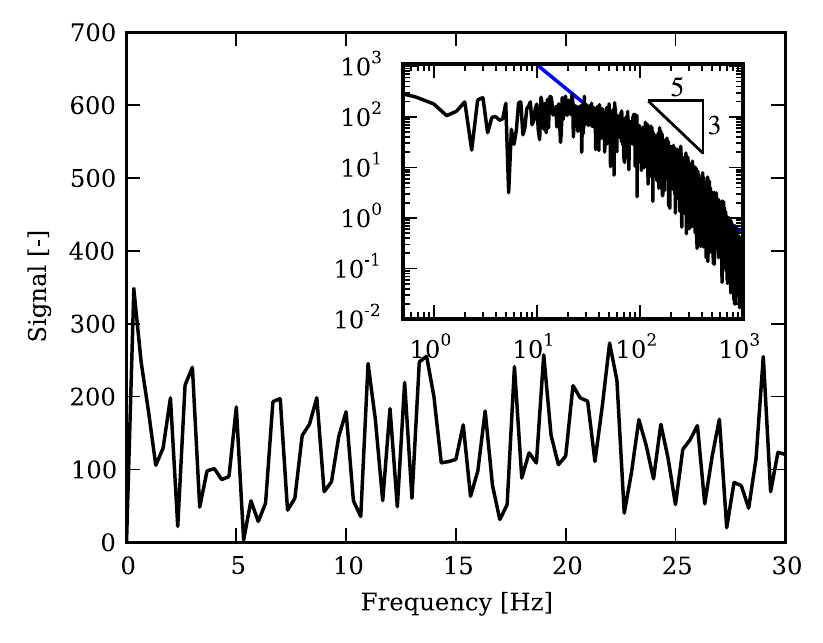}\hfill{}

}\protect\caption{DFT of temperature and velocity magnitude at monitoring point x=\SI{1}{\milli\meter},
y=\SI{0}{\milli\meter}, z=\SI{1}{\milli\meter}. Inserts on log-log
scale, in which, as a reference, a blue line with -5/3 slope is shown.\label{fig:3D-Velocity-magnitude-monitoring}}
\end{figure}

The frequency spectrum of the temperature and velocity magnitude signals
at the monitoring point, obtained by a discrete Fourier transform
(DFT) of the signals for the quasi steady state period between 2 and
5 seconds after the onset of heating, is shown in figures~\ref{fig:3D-temperature-probe-fft}
and \ref{fig:3D-velmag-probe-fft}, respectively. Both spectra exhibit
multiple peaks in the low-frequency region up to \SI{10}{\hertz}.
Due to the low Prandtl number of the fluid, momentum diffusivity is
small compared to thermal diffusivity. As a result, high frequency
oscillations are more strongly damped for temperature as compared
to velocity. The temperature signal drops beyond \SI{10}{\hertz},
whereas the velocity signal only starts dropping around \SI{30}{\hertz}.
The most dominant frequencies in the temperature spectrum are around
5, 7 and 11 Hertz, which also appear in the spectrum of the velocity
magnitude, though accompanied here by many other peaks up to 30 Hertz.

\subsection{The 3D nature of the flow instabilities}

\begin{figure*}
\includegraphics[width=7in]{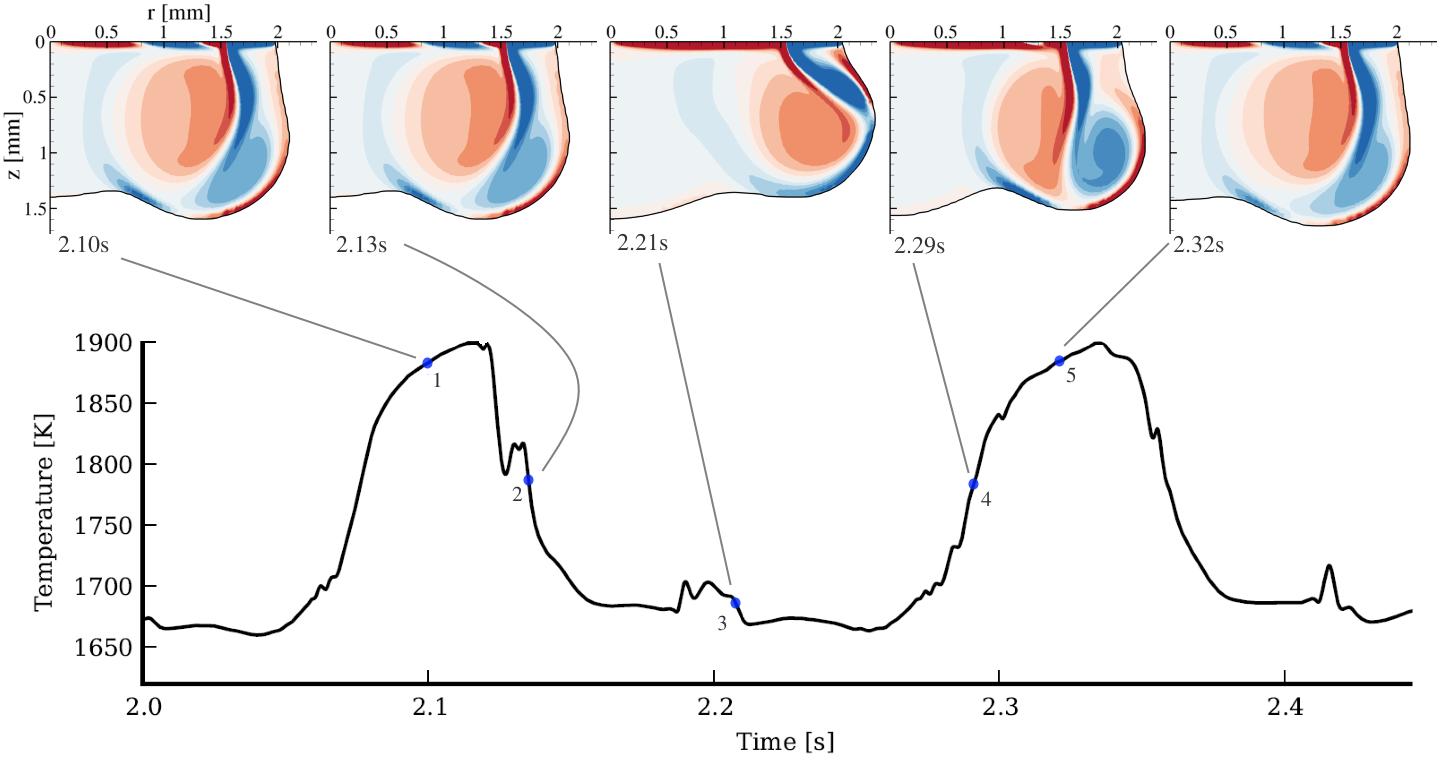}

\protect\caption{Two-dimensional flow vorticity at various time instances in the axisymmetric
case, with corresponding temperature probe at a monitoring point at
r=1mm, z=1mm\label{fig:2Dflow-and-temperature}}

\end{figure*}

To unravel to which extent the complexity and oscillating instability
of the melt pool flow is related to its three-dimensionality, we have
also performed a high-fidelity two-dimensional axisymmetric simulation
with a mesh that was very similar to that of the 3D simulations, and
identical numerical schemes. The 2D simulated flow, unlike earlier
published 2D flow simulations with enhanced transport coefficients
on coarse meshes with diffusive numerical schemes, exhibits a highly
unstable nature, very similar to that observed in the 3D simulations.
A characteristic flow oscillation is shown in figure~\ref{fig:2Dflow-and-temperature}.
The, now axisymmetric, circular downward jet stemming from the stagnation
point at the free surface shows qualitatively the same oscillatory
pattern as observed in the 3D simulations. It may therefore be concluded
that the additional degree of freedom of three-dimensionality is not
a requirement for triggering the jet instability. However, in the
2D case the oscillation is much more regular, which can easily be
observed from in the temperature signal at the monitoring point (figure~\ref{fig:2Dflow-and-temperature}).
While the amplitude of the oscillation of roughly \SI{250}{\kelvin}
is even larger than in the 3D flow, the oscillation frequency is low
and regular, with large peaks in the spectrum reoccurring at roughly
\SI{4}{\hertz}.

A similar oscillation pattern has been reported in literature for
the case of a slot jet impinging on a concave curved wall \citep{Gilard2005Slot}.
However, in the presently studied melt pool flow, at least in 2D,
the instability appears not to be simply due to the interaction between
the impinging jet and the concave bottom of the pool, but stems from
a more complex interplay between the jet, its oscillating origin,
and the constantly deforming melt pool bottom due to melting and re-solidification.
We performed a 2D axisymmetric simulation in which, after a quasi
steady-state with strong jet oscillations had been reached, further
melting (and re-solidification) of the pool boundary was prevented
by artificially increasing the latent heat of melting by a factor
100, thus de facto fixing the pool boundary. With this fixed pool
boundary, the jet oscillations quickly died out and the flow became
steady-state. This demonstrates that, at least in 2D, the interplay
with a deforming melt pool boundary is necessary to sustain the jet
oscillations. The same fixation of the pool boundary in 3D did not
lead to a suppression of the oscillations, indicating that the 3D
case is more prone to instabilities than the 2D case.

\section{Conclusion and outlook}

We have simulated the liquid metal flow in the melt pool of a conduction-mode
laser weld, using high-fidelity direct numerical simulations to gain
insight into flow instabilities that have been reported in experimental
studies, but not in numerical studies to date. 

In our simulations, unlike earlier numerical studies of weld pool flows that used coarse meshes, diffusive numerical schemes and enhanced transport coefficients, we observe self-sustained flow instabilities. These arise even in the absence of a deformable liquid-gas surface, temperature dependent material properties, or non-uniform surfactant concentrations. The instabilities even occur when restricting the flow to axial symmetry, albeit the oscillation is much more regular in that case. 

The amplitude and frequency spectrum of the temperature and velocity
oscillations support the argument of enhanced heat and momentum transport
due to turbulent flow in the melt pool. Averaged in space, the turbulent
diffusivities are approximately seven times higher than their molecular
values. This turbulent transport enhancement is of the same order
of magnitude as ad-hoc enhancement factors commonly used in previous
studies to obtain agreement between numerical weld pool simulations
and experimentally observed weld pool shapes. However, the observed
turbulent enhancement is strongly non-uniform, and highest in the
regions of the oscillating jets near the rim of the weld pool. Therefore,
unlike in simulations assuming uniform transport enhancement and unlike
experimentally observed for this case, our simulations lead to W-shaped
melt pools that are deepest near the rim of the pool. More generally though, W-shaped pools have been observed by many authors \citep{Winkler2000Effect,Ehlen2003Simulation,Pavlyk2001Numerical,Aidun1997Effect,Daub2012Erhohung,Tanaka2005Effects}.

Thus, while we have clearly demonstrated the presence of turbulent
flow instabilities and turbulent transport enhancement in laser weld
pools, the predicted turbulence does not lead to proper melt pool
shape predictions. We believe the most likely deficiency of our model
to be in the assumed uniform surfactant concentration due to the lack
of a mass transport model for surfactant species. \citet{Winkler2000Effect}
have demonstrated that surfactant concentrations may actually be highly
non-uniform, leading to strong alterations of the Marangoni forces
and flow. The stagnating flow at the jet origin will lead to a local high concentration of surface active element \citep{Winkler2005Multicomponent}, strengthening the local surface tension maximum and thus amplifying the Marangoni forces and the resulting flow. Hence, combining the results from \citet{Winkler2000Effect} with the present results, direct numerical or dynamic large eddy turbulence simulations coupled with a mass transport model for the surfactants are probably needed to move forward towards really predictive melt pool models for welding.

\section*{Acknowledgments}

We would like to thank the European Commission for funding the MINTWELD
project (reference 229108) via the FP7-NMP program. We thank SURFsara
for the support in using the Lisa Compute Cluster (project MP-235-12).

\bibliographystyle{elsarticle-num-names}
\bibliography{ijThermSci}

\begin{thebibliography}{61}
\providecommand{\natexlab}[1]{#1}
\providecommand{\url}[1]{\texttt{#1}}
\providecommand{\urlprefix}{URL }
\expandafter\ifx\csname urlstyle\endcsname\relax
  \providecommand{\doi}[1]{doi:\discretionary{}{}{}#1}\else
  \providecommand{\doi}[1]{doi:\discretionary{}{}{}\begingroup
  \urlstyle{rm}\url{#1}\endgroup}\fi
\providecommand{\bibinfo}[2]{#2}

\bibitem[{Winkler et~al.(1997)Winkler, Amberg, Inoue, and
  Koseki}]{Winkler1997Numerical}
\bibinfo{author}{C.~Winkler}, \bibinfo{author}{G.~Amberg},
  \bibinfo{author}{H.~Inoue}, \bibinfo{author}{T.~Koseki}, \bibinfo{title}{{A
  numerical and experimental investigation of qualitatively different weld pool
  shapes}}, in: \bibinfo{editor}{H.~Cerjak}, \bibinfo{editor}{H.~K. D.~H.
  Bhadeshia} (Eds.), \bibinfo{booktitle}{Mathematical Modelling of Weld
  Phenomena 4}, Materials Modelling Series, \bibinfo{publisher}{IOM
  Communications Ltd}, \bibinfo{address}{London}, ISBN
  \bibinfo{isbn}{1-86125-060-6}, \bibinfo{pages}{37--69}, \bibinfo{year}{1997}.

\bibitem[{Pavlyk and Dilthey(2001)}]{Pavlyk2001Numerical}
\bibinfo{author}{V.~Pavlyk}, \bibinfo{author}{U.~Dilthey}, \bibinfo{title}{{A
  numerical and experimental study of fluid flow and heat transfer in
  stationary GTA weld pools}}, in: \bibinfo{editor}{H.~Cerjak},
  \bibinfo{editor}{H.~K. D.~H. Bhadeshia} (Eds.),
  \bibinfo{booktitle}{Mathematical Modelling of Weld Phenomena 5}, Materials
  Modelling Series, \bibinfo{publisher}{IOM Communications Ltd}, ISBN
  \bibinfo{isbn}{1 86125 115 7}, \bibinfo{pages}{135--163},
  \bibinfo{year}{2001}.

\bibitem[{Anderson et~al.(2010)Anderson, DuPont, and
  DebRoy}]{Anderson2010Origin}
\bibinfo{author}{T.~D. Anderson}, \bibinfo{author}{J.~N. DuPont},
  \bibinfo{author}{T.~DebRoy}, \bibinfo{title}{{Origin of stray grain formation
  in single-crystal superalloy weld pools from heat transfer and fluid flow
  modeling}}, \bibinfo{journal}{Acta Materialia}
  \bibinfo{volume}{58}~(\bibinfo{number}{4}) (\bibinfo{year}{2010})
  \bibinfo{pages}{1441--1454}, ISSN \bibinfo{issn}{13596454},
  \doi{\bibinfo{doi}{10.1016/j.actamat.2009.10.051}}.

\bibitem[{De and DebRoy(2003)}]{De2003Probing}
\bibinfo{author}{A.~De}, \bibinfo{author}{T.~DebRoy}, \bibinfo{title}{{Probing
  unknown welding parameters from convective heat transfer calculation and
  multivariable optimization}}, \bibinfo{journal}{Journal of Physics D: Applied
  Physics} \bibinfo{volume}{37}~(\bibinfo{number}{1}) (\bibinfo{year}{2003})
  \bibinfo{pages}{140+}, ISSN \bibinfo{issn}{0022-3727},
  \doi{\bibinfo{doi}{10.1088/0022-3727/37/1/023}}.

\bibitem[{De and DebRoy(2004)}]{De2004Smart}
\bibinfo{author}{A.~De}, \bibinfo{author}{T.~DebRoy}, \bibinfo{title}{{A smart
  model to estimate effective thermal conductivity and viscosity in the weld
  pool}}, \bibinfo{journal}{Journal of Applied Physics}
  \bibinfo{volume}{95}~(\bibinfo{number}{9}) (\bibinfo{year}{2004})
  \bibinfo{pages}{5230--5240}, \doi{\bibinfo{doi}{10.1063/1.1695593}}.

\bibitem[{De and DebRoy(2006)}]{De2006Improving}
\bibinfo{author}{A.~De}, \bibinfo{author}{T.~DebRoy},
  \bibinfo{title}{{Improving reliability of heat and fluid flow calculation
  during conduction mode laser spot welding by multivariable optimisation}},
  \bibinfo{journal}{Science and Technology of Welding and Joining}
  \bibinfo{volume}{11}~(\bibinfo{number}{2}) (\bibinfo{year}{2006})
  \bibinfo{pages}{143--153}, ISSN \bibinfo{issn}{1362-1718},
  \doi{\bibinfo{doi}{10.1179/174329306x84346}}.

\bibitem[{Pitscheneder et~al.(1996)Pitscheneder, DebRoy, Mundra, and
  Ebner}]{Pitscheneder1996Role}
\bibinfo{author}{W.~Pitscheneder}, \bibinfo{author}{T.~DebRoy},
  \bibinfo{author}{K.~Mundra}, \bibinfo{author}{R.~Ebner},
  \bibinfo{title}{{Role of sulfur and processing variables on the temporal
  evolution of weld pool geometry during multikilowatt laser beam welding of
  steels}}, \bibinfo{journal}{Welding Journal}
  \bibinfo{volume}{75}~(\bibinfo{number}{3}) (\bibinfo{year}{1996})
  \bibinfo{pages}{71--s--80--s}.

\bibitem[{Tan et~al.(2012)Tan, Bailey, and Shin}]{Tan2012Numerical}
\bibinfo{author}{W.~Tan}, \bibinfo{author}{N.~S. Bailey},
  \bibinfo{author}{Y.~C. Shin}, \bibinfo{title}{{Numerical Modeling of
  Transport Phenomena and Dendritic Growth in Laser Spot Conduction Welding of
  304 Stainless Steel}}, \bibinfo{journal}{Journal of Manufacturing Science and
  Engineering} \bibinfo{volume}{134}~(\bibinfo{number}{4})
  (\bibinfo{year}{2012}) \bibinfo{pages}{041010+}, ISSN
  \bibinfo{issn}{10871357}, \doi{\bibinfo{doi}{10.1115/1.4007101}}.

\bibitem[{Mishra et~al.(2008)Mishra, Lienert, Johnson, and
  DebRoy}]{Mishra2008Experimental}
\bibinfo{author}{S.~Mishra}, \bibinfo{author}{T.~J. Lienert},
  \bibinfo{author}{M.~Q. Johnson}, \bibinfo{author}{T.~DebRoy},
  \bibinfo{title}{{An experimental and theoretical study of gas tungsten arc
  welding of stainless steel plates with different sulfur concentrations}},
  \bibinfo{journal}{Acta Materialia} \bibinfo{volume}{56}~(\bibinfo{number}{9})
  (\bibinfo{year}{2008}) \bibinfo{pages}{2133--2146}, ISSN
  \bibinfo{issn}{13596454}, \doi{\bibinfo{doi}{10.1016/j.actamat.2008.01.028}}.

\bibitem[{Kraus(1989)}]{Kraus1989Surface}
\bibinfo{author}{H.~G. Kraus}, \bibinfo{title}{{Surface Temperature
  Measurements of GTA Weld Pools on Thin-Plate 304 Stainless Steel}},
  \bibinfo{journal}{Welding Journal} \bibinfo{volume}{68}~(\bibinfo{number}{3})
  (\bibinfo{year}{1989}) \bibinfo{pages}{84--s--91--s}.

\bibitem[{Zehr(1991)}]{Zehr1991Thermocapillary}
\bibinfo{author}{R.~L. Zehr}, \bibinfo{title}{{Thermocapillary convection in
  laser melted pools during materials processing}}, Ph.D. thesis,
  \bibinfo{school}{University of Illinois at Urbana-Champaign},
  \bibinfo{year}{1991}.

\bibitem[{Zhao et~al.(2009)Zhao, van Steijn, Richardson, Kleijn, Kenjeres, and
  Saldi}]{Zhao2009Unsteady}
\bibinfo{author}{C.~X. Zhao}, \bibinfo{author}{V.~van Steijn},
  \bibinfo{author}{I.~M. Richardson}, \bibinfo{author}{C.~R. Kleijn},
  \bibinfo{author}{S.~Kenjeres}, \bibinfo{author}{Z.~Saldi},
  \bibinfo{title}{{Unsteady interfacial phenomena during inward weld pool flow
  with an active surface oxide}}, \bibinfo{journal}{Science and Technology of
  Welding \& Joining} \bibinfo{volume}{14}~(\bibinfo{number}{2})
  (\bibinfo{year}{2009}) \bibinfo{pages}{132--140}, ISSN
  \bibinfo{issn}{1362-1718}, \doi{\bibinfo{doi}{10.1179/136217108x370281}}.

\bibitem[{Zhao(2011)}]{Zhao2011Measurements}
\bibinfo{author}{C.~Zhao}, \bibinfo{title}{{Measurements of fluid flow in weld
  pools}}, Ph.D. thesis, \bibinfo{school}{Delft University of Technology},
  \bibinfo{year}{2011}.

\bibitem[{Ehlen et~al.(2003)Ehlen, Ludwig, and Sahm}]{Ehlen2003Simulation}
\bibinfo{author}{G.~Ehlen}, \bibinfo{author}{A.~Ludwig}, \bibinfo{author}{P.~R.
  Sahm}, \bibinfo{title}{{Simulation of time-dependent pool shape during laser
  spot welding: Transient effects}}, \bibinfo{journal}{Metallurgical and
  Materials Transactions A} \bibinfo{volume}{34}~(\bibinfo{number}{12})
  (\bibinfo{year}{2003}) \bibinfo{pages}{2947--2961}, ISSN
  \bibinfo{issn}{1073-5623}, \doi{\bibinfo{doi}{10.1007/s11661-003-0194-x}}.

\bibitem[{Saldi et~al.(2013)Saldi, Kidess, Kenjere\v{s}, Zhao, Richardson, and
  Kleijn}]{Saldi2013Effect}
\bibinfo{author}{Z.~S. Saldi}, \bibinfo{author}{A.~Kidess},
  \bibinfo{author}{S.~Kenjere\v{s}}, \bibinfo{author}{C.~Zhao},
  \bibinfo{author}{I.~M. Richardson}, \bibinfo{author}{C.~R. Kleijn},
  \bibinfo{title}{{Effect of enhanced heat and mass transport and flow reversal
  during cool down on weld pool shapes in laser spot welding of steel}},
  \bibinfo{journal}{International Journal of Heat and Mass Transfer}
  \bibinfo{volume}{66} (\bibinfo{year}{2013}) \bibinfo{pages}{879--888}, ISSN
  \bibinfo{issn}{00179310},
  \doi{\bibinfo{doi}{10.1016/j.ijheatmasstransfer.2013.07.085}}.

\bibitem[{Ha and Kim(2005)}]{Ha2005Study}
\bibinfo{author}{E.-J. Ha}, \bibinfo{author}{W.-S. Kim}, \bibinfo{title}{{A
  study of low-power density laser welding process with evolution of free
  surface}}, \bibinfo{journal}{International Journal of Heat and Fluid Flow}
  \bibinfo{volume}{26}~(\bibinfo{number}{4}) (\bibinfo{year}{2005})
  \bibinfo{pages}{613--621}, ISSN \bibinfo{issn}{0142727X},
  \doi{\bibinfo{doi}{10.1016/j.ijheatfluidflow.2005.03.009}}.

\bibitem[{Winkler et~al.(2000)Winkler, Amberg, Inoue, Koseki, and
  Fuji}]{Winkler2000Effect}
\bibinfo{author}{C.~Winkler}, \bibinfo{author}{G.~Amberg},
  \bibinfo{author}{H.~Inoue}, \bibinfo{author}{T.~Koseki},
  \bibinfo{author}{M.~Fuji}, \bibinfo{title}{Effect of surfactant
  redistribution on weld pool shape during gas tungsten arc welding},
  \bibinfo{journal}{Science and Technology of Welding and Joining}
  \bibinfo{volume}{5}~(\bibinfo{number}{1}) (\bibinfo{year}{2000})
  \bibinfo{pages}{8--20}, ISSN \bibinfo{issn}{1362-1718},
  \doi{\bibinfo{doi}{10.1179/stw.2000.5.1.8}}.

\bibitem[{Winkler and Amberg(2005)}]{Winkler2005Multicomponent}
\bibinfo{author}{C.~Winkler}, \bibinfo{author}{G.~Amberg},
  \bibinfo{title}{Multicomponent surfactant mass transfer in GTA-welding},
  \bibinfo{journal}{Progress in Computational Fluid Dynamics, An International
  Journal} \bibinfo{volume}{5}~(\bibinfo{number}{3-5}) (\bibinfo{year}{2005})
  \bibinfo{pages}{190--206}, ISSN \bibinfo{issn}{1468-4349},
  \doi{\bibinfo{doi}{10.1504/pcfd.2005.006754}}.

\bibitem[{Sahoo et~al.(1988)Sahoo, Debroy, and McNallan}]{Sahoo1988Surface}
\bibinfo{author}{P.~Sahoo}, \bibinfo{author}{T.~Debroy},
  \bibinfo{author}{M.~McNallan}, \bibinfo{title}{{Surface tension of binary
  metal - surface active solute systems under conditions relevant to welding
  metallurgy}}, \bibinfo{journal}{Metallurgical and Materials Transactions B}
  \bibinfo{volume}{19}~(\bibinfo{number}{3}) (\bibinfo{year}{1988})
  \bibinfo{pages}{483--491--491}, ISSN \bibinfo{issn}{0360-2141},
  \doi{\bibinfo{doi}{10.1007/bf02657748}}.

\bibitem[{He et~al.(2005)He, Elmer, and DebRoy}]{He2005Heat}
\bibinfo{author}{X.~He}, \bibinfo{author}{J.~W. Elmer},
  \bibinfo{author}{T.~DebRoy}, \bibinfo{title}{Heat transfer and fluid flow in
  laser microwelding}, \bibinfo{journal}{Journal of Applied Physics}
  \bibinfo{volume}{97}~(\bibinfo{number}{8}) (\bibinfo{year}{2005})
  \bibinfo{pages}{084909+}, ISSN \bibinfo{issn}{0021-8979},
  \doi{\bibinfo{doi}{10.1063/1.1873032}}.

\bibitem[{Roy et~al.(2006)Roy, Elmer, and DebRoy}]{Roy2006Mathematical}
\bibinfo{author}{G.~G. Roy}, \bibinfo{author}{J.~W. Elmer},
  \bibinfo{author}{T.~DebRoy}, \bibinfo{title}{Mathematical modeling of heat
  transfer, fluid flow, and solidification during linear welding with a pulsed
  laser beam}, \bibinfo{journal}{Journal of Applied Physics}
  \bibinfo{volume}{100}~(\bibinfo{number}{3}) (\bibinfo{year}{2006})
  \bibinfo{pages}{034903+}, \doi{\bibinfo{doi}{10.1063/1.2214392}}.

\bibitem[{Choo and Szekely(1994)}]{Choo1994Possible}
\bibinfo{author}{R.~T.~C. Choo}, \bibinfo{author}{J.~Szekely},
  \bibinfo{title}{{The Possible Role of Turbulence in GTA Weld Pool
  Behaviour}}, \bibinfo{journal}{Welding Journal}
  \bibinfo{volume}{73}~(\bibinfo{number}{2}).

\bibitem[{Chakraborty et~al.(2003)Chakraborty, Chakraborty, and
  Dutta}]{Chakraborty2003Modelling}
\bibinfo{author}{N.~Chakraborty}, \bibinfo{author}{S.~Chakraborty},
  \bibinfo{author}{P.~Dutta}, \bibinfo{title}{{Modelling of turbulent transport
  in arc welding pools}}, \bibinfo{journal}{International Journal of Numerical
  Methods for Heat \& Fluid Flow} \bibinfo{volume}{13}~(\bibinfo{number}{1})
  (\bibinfo{year}{2003}) \bibinfo{pages}{7--30}, ISSN
  \bibinfo{issn}{0961-5539}, \doi{\bibinfo{doi}{10.1108/09615530310456741}}.

\bibitem[{Chakraborty et~al.(2004{\natexlab{a}})Chakraborty, Chatterjee, and
  Chakraborty}]{Chakraborty2004MODELING}
\bibinfo{author}{N.~Chakraborty}, \bibinfo{author}{D.~Chatterjee},
  \bibinfo{author}{S.~Chakraborty}, \bibinfo{title}{Modeling of turbulent
  transport in laser surface alloying}, \bibinfo{journal}{Numerical Heat
  Transfer, Part A: Applications} \bibinfo{volume}{46}~(\bibinfo{number}{10})
  (\bibinfo{year}{2004}{\natexlab{a}}) \bibinfo{pages}{1009--1032},
  \doi{\bibinfo{doi}{10.1080/10407780490517629}}.

\bibitem[{Chakraborty and Chakraborty(2005)}]{Chakraborty2005Influences}
\bibinfo{author}{N.~Chakraborty}, \bibinfo{author}{S.~Chakraborty},
  \bibinfo{title}{{Influences of Sign of Surface Tension Coefficient on
  Turbulent Weld Pool Convection in a Gas Tungsten Arc Welding (GTAW) Process:
  A Comparative Study}}, \bibinfo{journal}{Journal of Heat Transfer}
  \bibinfo{volume}{127}~(\bibinfo{number}{8}) (\bibinfo{year}{2005})
  \bibinfo{pages}{848+}, ISSN \bibinfo{issn}{00221481},
  \doi{\bibinfo{doi}{10.1115/1.1928913}}.

\bibitem[{Chakraborty et~al.(2004{\natexlab{b}})Chakraborty, Chakraborty, and
  Dutta}]{Chakraborty2004THREEDIMENSIONAL}
\bibinfo{author}{N.~Chakraborty}, \bibinfo{author}{S.~Chakraborty},
  \bibinfo{author}{P.~Dutta}, \bibinfo{title}{Three-dimensional modeling of
  turbulent weld pool convection in GTAW processes},
  \bibinfo{journal}{Numerical Heat Transfer, Part A: Applications}
  \bibinfo{volume}{45}~(\bibinfo{number}{4})
  (\bibinfo{year}{2004}{\natexlab{b}}) \bibinfo{pages}{391--413},
  \doi{\bibinfo{doi}{10.1080/10407780490250364}}.

\bibitem[{Chakraborty and
  Chakraborty(2007{\natexlab{a}})}]{Chakraborty2007Modelling}
\bibinfo{author}{N.~Chakraborty}, \bibinfo{author}{S.~Chakraborty},
  \bibinfo{title}{{Modelling of turbulent molten pool convection in laser
  welding of a copper-nickel dissimilar couple}},
  \bibinfo{journal}{International Journal of Heat and Mass Transfer}
  \bibinfo{volume}{50}~(\bibinfo{number}{9-10})
  (\bibinfo{year}{2007}{\natexlab{a}}) \bibinfo{pages}{1805--1822}, ISSN
  \bibinfo{issn}{00179310},
  \doi{\bibinfo{doi}{10.1016/j.ijheatmasstransfer.2006.10.030}}.

\bibitem[{Dong et~al.(2011)Dong, Lu, Li, and Li}]{Dong2011GTAW}
\bibinfo{author}{W.~Dong}, \bibinfo{author}{S.~Lu}, \bibinfo{author}{D.~Li},
  \bibinfo{author}{Y.~Li}, \bibinfo{title}{{GTAW liquid pool convections and
  the weld shape variations under helium gas shielding}},
  \bibinfo{journal}{International Journal of Heat and Mass Transfer}
  \bibinfo{volume}{54}~(\bibinfo{number}{7-8}) (\bibinfo{year}{2011})
  \bibinfo{pages}{1420--1431}, ISSN \bibinfo{issn}{00179310},
  \doi{\bibinfo{doi}{10.1016/j.ijheatmasstransfer.2010.07.069}}.

\bibitem[{Goodarzi et~al.(1998)Goodarzi, Choo, Takasu, and
  Toguri}]{Goodarzi1998Effect}
\bibinfo{author}{M.~Goodarzi}, \bibinfo{author}{R.~Choo},
  \bibinfo{author}{T.~Takasu}, \bibinfo{author}{J.~M. Toguri},
  \bibinfo{title}{{The effect of the cathode tip angle on the gas tungsten arc
  welding arc and weld pool: II. The mathematical model for the weld pool}},
  \bibinfo{journal}{Journal of Physics D: Applied Physics}
  \bibinfo{volume}{31}~(\bibinfo{number}{5}) (\bibinfo{year}{1998})
  \bibinfo{pages}{569+}, ISSN \bibinfo{issn}{0022-3727},
  \doi{\bibinfo{doi}{10.1088/0022-3727/31/5/014}}.

\bibitem[{Hong et~al.(1998)Hong, Weckman, and Strong}]{Hong1998Influence}
\bibinfo{author}{K.~Hong}, \bibinfo{author}{D.~Weckman},
  \bibinfo{author}{A.~Strong}, \bibinfo{title}{{The influence of thermofluids
  phenomena in gas tungsten arc welds in high and low thermal conductivity
  metals}}, \bibinfo{journal}{Canadian Metallurgical Quarterly}
  \bibinfo{volume}{37}~(\bibinfo{number}{3-4}) (\bibinfo{year}{1998})
  \bibinfo{pages}{293--303}, ISSN \bibinfo{issn}{00084433},
  \doi{\bibinfo{doi}{10.1016/s0008-4433(97)00021-9}}.

\bibitem[{Hong et~al.(2002)Hong, Weckman, Strong, and
  Zheng}]{Hong2002Modelling}
\bibinfo{author}{K.~Hong}, \bibinfo{author}{D.~C. Weckman},
  \bibinfo{author}{A.~B. Strong}, \bibinfo{author}{W.~Zheng},
  \bibinfo{title}{{Modelling turbulent thermofluid flow in stationary gas
  tungsten arc weld pools}}, \bibinfo{journal}{Science and Technology of
  Welding and Joining}  (\bibinfo{year}{2002}) \bibinfo{pages}{125--136}ISSN
  \bibinfo{issn}{1362-1718}, \doi{\bibinfo{doi}{10.1179/136217102225002619}}.

\bibitem[{Hong et~al.(2003)Hong, Weckman, Strong, and
  Zheng}]{Hong2003Vorticity}
\bibinfo{author}{K.~Hong}, \bibinfo{author}{D.~C. Weckman},
  \bibinfo{author}{A.~B. Strong}, \bibinfo{author}{W.~Zheng},
  \bibinfo{title}{{Vorticity based turbulence model for thermofluids modelling
  of welds}}, \bibinfo{journal}{Science and Technology of Welding and Joining}
  (\bibinfo{year}{2003}) \bibinfo{pages}{313--324}ISSN
  \bibinfo{issn}{1362-1718}, \doi{\bibinfo{doi}{10.1179/136217103225005507}}.

\bibitem[{Jaidi et~al.(2002)Jaidi, Murthy, and Dutta}]{Jaidi2002kEps}
\bibinfo{author}{J.~Jaidi}, \bibinfo{author}{K.~S.~S. Murthy},
  \bibinfo{author}{P.~Dutta}, \bibinfo{title}{{A k-{$\epsilon$} Model for
  Turbulent Weld Pool Convection in Gas Metal Arc Welding Process}}, in:
  \bibinfo{editor}{S.~A. David} (Ed.), \bibinfo{booktitle}{6th International
  Trends in Welding Research Conference Proceedings}, Trends in Welding
  Research, \bibinfo{organization}{ASM International}, \bibinfo{publisher}{ASM
  International}, \bibinfo{pages}{147--152}, \bibinfo{year}{2002}.

\bibitem[{Jaidi and Dutta(2004)}]{Jaidi2004Threedimensional}
\bibinfo{author}{J.~Jaidi}, \bibinfo{author}{P.~Dutta},
  \bibinfo{title}{{Three-dimensional turbulent weld pool convection in gas
  metal arc welding process}}, \bibinfo{journal}{Science and Technology of
  Welding and Joining} \bibinfo{volume}{9}~(\bibinfo{number}{5})
  (\bibinfo{year}{2004}) \bibinfo{pages}{407--414}, ISSN
  \bibinfo{issn}{1362-1718}, \doi{\bibinfo{doi}{10.1179/136217104225021814}}.

\bibitem[{Skouras et~al.(2010)Skouras, Chakraborty, and
  Chakraborty}]{Skouras2010Computational}
\bibinfo{author}{A.~K. Skouras}, \bibinfo{author}{N.~Chakraborty},
  \bibinfo{author}{S.~Chakraborty}, \bibinfo{title}{{Computational Analysis of
  the Effects of Process Parameters on Molten Pool Transport in Cu-Ni
  Dissimilar Laser Weld Pool}}, \bibinfo{journal}{Numerical Heat Transfer, Part
  A: Applications} \bibinfo{volume}{58}~(\bibinfo{number}{4})
  (\bibinfo{year}{2010}) \bibinfo{pages}{272--294},
  \doi{\bibinfo{doi}{10.1080/10407782.2010.505154}}.

\bibitem[{Wang et~al.(2014)Wang, Fan, Huang, and Huang}]{Wang2014Unified}
\bibinfo{author}{X.~Wang}, \bibinfo{author}{D.~Fan},
  \bibinfo{author}{J.~Huang}, \bibinfo{author}{Y.~Huang}, \bibinfo{title}{A
  unified model of coupled arc plasma and weld pool for double electrodes TIG
  welding}, \bibinfo{journal}{Journal of Physics D: Applied Physics}
  \bibinfo{volume}{47}~(\bibinfo{number}{27}) (\bibinfo{year}{2014})
  \bibinfo{pages}{275202+}, ISSN \bibinfo{issn}{0022-3727},
  \doi{\bibinfo{doi}{10.1088/0022-3727/47/27/275202}}.

\bibitem[{Chatterjee and Chakraborty(2005)}]{Chatterjee2005Largeeddy}
\bibinfo{author}{D.~Chatterjee}, \bibinfo{author}{S.~Chakraborty},
  \bibinfo{title}{{Large-eddy simulation of laser-induced
  surface-tension-driven flow}}, \bibinfo{journal}{Metallurgical and Materials
  Transactions B} \bibinfo{volume}{36}~(\bibinfo{number}{6})
  (\bibinfo{year}{2005}) \bibinfo{pages}{743--754}, ISSN
  \bibinfo{issn}{1073-5615}, \doi{\bibinfo{doi}{10.1007/s11663-005-0078-0}}.

\bibitem[{Chakraborty and
  Chakraborty(2007{\natexlab{b}})}]{Chakraborty2007Thermal2}
\bibinfo{author}{N.~Chakraborty}, \bibinfo{author}{S.~Chakraborty},
  \bibinfo{title}{Thermal Transport Regimes and Generalized Regime Diagram for
  High Energy Surface Melting Processes}, \bibinfo{journal}{Metallurgical and
  Materials Transactions B} \bibinfo{volume}{38}~(\bibinfo{number}{1})
  (\bibinfo{year}{2007}{\natexlab{b}}) \bibinfo{pages}{143--147}, ISSN
  \bibinfo{issn}{1073-5615}, \doi{\bibinfo{doi}{10.1007/s11663-006-9000-7}}.

\bibitem[{Chakraborty(2007)}]{Chakraborty2007Thermal}
\bibinfo{author}{N.~Chakraborty}, \bibinfo{title}{{Thermal Transport Regimes
  and Effects of Prandtl Number in Molten Pool Transport in Laser Surface
  Melting Processes}}, \bibinfo{journal}{Numerical Heat Transfer, Part A:
  Applications} \bibinfo{volume}{53}~(\bibinfo{number}{3})
  (\bibinfo{year}{2007}) \bibinfo{pages}{273--294},
  \doi{\bibinfo{doi}{10.1080/10407780701557709}}.

\bibitem[{Singh et~al.(2001)Singh, Pardeshi, and Basu}]{Singh2001Modelling}
\bibinfo{author}{A.~Singh}, \bibinfo{author}{R.~Pardeshi},
  \bibinfo{author}{B.~Basu}, \bibinfo{title}{{Modelling of convection during
  solidification of metal and alloys}}, \bibinfo{journal}{Sadhana}
  \bibinfo{volume}{26}~(\bibinfo{number}{1}) (\bibinfo{year}{2001})
  \bibinfo{pages}{139--162}, ISSN \bibinfo{issn}{0256-2499},
  \doi{\bibinfo{doi}{10.1007/bf02728483}}.

\bibitem[{Breugem et~al.(2006)Breugem, Boersma, and
  Uittenbogaard}]{Breugem2006Influence}
\bibinfo{author}{W.~P. Breugem}, \bibinfo{author}{B.~J. Boersma},
  \bibinfo{author}{R.~E. Uittenbogaard}, \bibinfo{title}{The influence of wall
  permeability on turbulent channel flow}, \bibinfo{journal}{Journal of Fluid
  Mechanics} \bibinfo{volume}{562} (\bibinfo{year}{2006})
  \bibinfo{pages}{35--72}, ISSN \bibinfo{issn}{1469-7645},
  \doi{\bibinfo{doi}{10.1017/s0022112006000887}}.

\bibitem[{Mills and Keene(1990)}]{Mills1990Factors}
\bibinfo{author}{K.~C. Mills}, \bibinfo{author}{B.~J. Keene},
  \bibinfo{title}{Factors affecting variable weld penetration},
  \bibinfo{journal}{International Materials Reviews}  (\bibinfo{year}{1990})
  \bibinfo{pages}{185--216}ISSN \bibinfo{issn}{0950-6608}.

\bibitem[{Ozawa et~al.(2014{\natexlab{a}})Ozawa, Morohoshi, and
  Hibiya}]{Ozawa2014InfluenceS}
\bibinfo{author}{S.~Ozawa}, \bibinfo{author}{K.~Morohoshi},
  \bibinfo{author}{T.~Hibiya}, \bibinfo{title}{Influence of Oxygen Partial
  Pressure on Surface Tension of Molten Type 304 and 316 Stainless Steels
  Measured by Oscillating Droplet Method Using Electromagnetic Levitation},
  \bibinfo{journal}{ISIJ International}
  \bibinfo{volume}{54}~(\bibinfo{number}{9})
  (\bibinfo{year}{2014}{\natexlab{a}}) \bibinfo{pages}{2097--2103}, ISSN
  \bibinfo{issn}{0915-1559},
  \doi{\bibinfo{doi}{10.2355/isijinternational.54.2097}}.

\bibitem[{Hibiya and Ozawa(2013)}]{Hibiya2013Effect}
\bibinfo{author}{T.~Hibiya}, \bibinfo{author}{S.~Ozawa}, \bibinfo{title}{Effect
  of oxygen partial pressure on the Marangoni flow of molten metals},
  \bibinfo{journal}{Cryst. Res. Technol.}
  \bibinfo{volume}{48}~(\bibinfo{number}{4}) (\bibinfo{year}{2013})
  \bibinfo{pages}{208--213}, \doi{\bibinfo{doi}{10.1002/crat.201200514}}.

\bibitem[{Ozawa et~al.(2014{\natexlab{b}})Ozawa, Takahashi, Watanabe, and
  Fukuyama}]{Ozawa2014Influence}
\bibinfo{author}{S.~Ozawa}, \bibinfo{author}{S.~Takahashi},
  \bibinfo{author}{N.~Watanabe}, \bibinfo{author}{H.~Fukuyama},
  \bibinfo{title}{Influence of Oxygen Adsorption on Surface Tension of Molten
  Nickel Measured Under Reducing Gas Atmosphere}
  \bibinfo{volume}{35}~(\bibinfo{number}{9-10})
  (\bibinfo{year}{2014}{\natexlab{b}}) \bibinfo{pages}{1705--1711},
  \doi{\bibinfo{doi}{10.1007/s10765-014-1674-5}}.

\bibitem[{Kou et~al.(2011)Kou, Limmaneevichitr, and Wei}]{kou2011oscillatory}
\bibinfo{author}{S.~Kou}, \bibinfo{author}{C.~Limmaneevichitr},
  \bibinfo{author}{P.~Wei}, \bibinfo{title}{Oscillatory Marangoni Flow: A
  Fundamental Study by Conduction-Mode Laser Spot Welding: Through Marangoni
  flow, a surface-active agent can affect not only the weld pool depth, but
  also the pool surface deformation, pool surface oscillation, and ripple
  formation}, \bibinfo{journal}{Welding Journal}
  \bibinfo{volume}{90}~(\bibinfo{number}{12}).

\bibitem[{Weller et~al.(1998)Weller, Tabor, Jasak, and
  Fureby}]{Weller1998Tensorial}
\bibinfo{author}{H.~G. Weller}, \bibinfo{author}{G.~Tabor},
  \bibinfo{author}{H.~Jasak}, \bibinfo{author}{C.~Fureby}, \bibinfo{title}{{A
  tensorial approach to computational continuum mechanics using object-oriented
  techniques}}, \bibinfo{journal}{Computers in Physics}
  \bibinfo{volume}{12}~(\bibinfo{number}{6}) (\bibinfo{year}{1998})
  \bibinfo{pages}{620--631}, \doi{\bibinfo{doi}{10.1063/1.168744}}.

\bibitem[{Berberovic(2010)}]{Berberovic2010Investigation}
\bibinfo{author}{E.~Berberovic}, \bibinfo{title}{{Investigation of Free-surface
  Flow Associated with Drop Impact: Numerical Simulations and Theoretical
  Modeling}}, Ph.D. thesis, \bibinfo{school}{Technische Universit{ae}t
  Darmstadt}, \bibinfo{year}{2010}.

\bibitem[{Issa(1986)}]{Issa1986Solution}
\bibinfo{author}{R.~I. Issa}, \bibinfo{title}{{Solution of the implicitly
  discretised fluid flow equations by operator-splitting}},
  \bibinfo{journal}{Journal of Computational Physics}
  \bibinfo{volume}{62}~(\bibinfo{number}{1}) (\bibinfo{year}{1986})
  \bibinfo{pages}{40--65}, ISSN \bibinfo{issn}{00219991},
  \doi{\bibinfo{doi}{10.1016/0021-9991(86)90099-9}}.

\bibitem[{Voller and Swaminathan(1991)}]{Voller1991GENERAL}
\bibinfo{author}{V.~R. Voller}, \bibinfo{author}{C.~R. Swaminathan},
  \bibinfo{title}{General source-based method for solidification phase change},
  \bibinfo{journal}{Numerical Heat Transfer, Part B: Fundamentals: An
  International Journal of Computation and Methodology}
  \bibinfo{volume}{19}~(\bibinfo{number}{2}) (\bibinfo{year}{1991})
  \bibinfo{pages}{175--189}, \doi{\bibinfo{doi}{10.1080/10407799108944962}}.

\bibitem[{Mills et~al.(1998)Mills, Keene, Brooks, and
  Shirali}]{Mills1998Marangoni}
\bibinfo{author}{K.~C. Mills}, \bibinfo{author}{B.~J. Keene},
  \bibinfo{author}{R.~F. Brooks}, \bibinfo{author}{A.~Shirali},
  \bibinfo{title}{Marangoni effects in welding},
  \bibinfo{journal}{Mathematical, Physical and Engineering Sciences}
  \bibinfo{volume}{356}~(\bibinfo{number}{1739}) (\bibinfo{year}{1998})
  \bibinfo{pages}{911--925}, ISSN \bibinfo{issn}{1471-2962},
  \doi{\bibinfo{doi}{10.1098/rsta.1998.0196}}.

\bibitem[{Keene et~al.(1982)Keene, Mills, Bryant, and
  Hondros}]{Keene1982Effects}
\bibinfo{author}{B.~J. Keene}, \bibinfo{author}{K.~C. Mills},
  \bibinfo{author}{J.~W. Bryant}, \bibinfo{author}{E.~D. Hondros},
  \bibinfo{title}{Effects of Interaction Between Surface Active Elements on the
  Surface Tension of Iron}, \bibinfo{journal}{Canadian Metallurgical Quarterly}
   (\bibinfo{year}{1982}) \bibinfo{pages}{393--403}ISSN
  \bibinfo{issn}{0008-4433}, \doi{\bibinfo{doi}{10.1179/000844382795243461}}.

\bibitem[{Nemchinsky(1997)}]{Nemchinsky1997Role}
\bibinfo{author}{V.~A. Nemchinsky}, \bibinfo{title}{The role of thermocapillary
  instability in heat transfer in a liquid metal pool},
  \bibinfo{journal}{International Journal of Heat and Mass Transfer}
  \bibinfo{volume}{40}~(\bibinfo{number}{4}) (\bibinfo{year}{1997})
  \bibinfo{pages}{881--891}, ISSN \bibinfo{issn}{00179310},
  \doi{\bibinfo{doi}{10.1016/0017-9310(96)00163-9}}.

\bibitem[{Kuhlmann and Schoisswohl(2010)}]{Kuhlmann2010Flow}
\bibinfo{author}{H.~C. Kuhlmann}, \bibinfo{author}{U.~Schoisswohl},
  \bibinfo{title}{Flow instabilities in thermocapillary-buoyant liquid pools},
  \bibinfo{journal}{Journal of Fluid Mechanics} \bibinfo{volume}{644}
  (\bibinfo{year}{2010}) \bibinfo{pages}{509--535}, ISSN
  \bibinfo{issn}{1469-7645}, \doi{\bibinfo{doi}{10.1017/s0022112009992953}}.

\bibitem[{Karcher et~al.(2000)Karcher, Schaller, Boeck, Metzner, and
  Thess}]{Karcher2000Turbulent}
\bibinfo{author}{C.~Karcher}, \bibinfo{author}{R.~Schaller},
  \bibinfo{author}{T.~Boeck}, \bibinfo{author}{C.~Metzner},
  \bibinfo{author}{A.~Thess}, \bibinfo{title}{Turbulent heat transfer in liquid
  iron during electron beam evaporation}, \bibinfo{journal}{International
  Journal of Heat and Mass Transfer}
  \bibinfo{volume}{43}~(\bibinfo{number}{10}) (\bibinfo{year}{2000})
  \bibinfo{pages}{1759--1766}, ISSN \bibinfo{issn}{00179310},
  \doi{\bibinfo{doi}{10.1016/s0017-9310(99)00248-3}}.

\bibitem[{Boeck and Karcher(2003)}]{Boeck2003LowPrandtlNumber}
\bibinfo{author}{T.~Boeck}, \bibinfo{author}{C.~Karcher},
  \bibinfo{title}{Low-Prandtl-Number Marangoni Convection Driven by Localized
  Heating on the Free Surface: Results of Three-Dimensional Direct
  Simulations}, in: \bibinfo{editor}{R.~Narayanan},
  \bibinfo{editor}{D.~Schwabe} (Eds.), \bibinfo{booktitle}{Interfacial Fluid
  Dynamics and Transport Processes}, vol. \bibinfo{volume}{628} of
  \emph{\bibinfo{series}{Lecture Notes in Physics}},
  \bibinfo{publisher}{Springer Berlin Heidelberg}, \bibinfo{pages}{157--175},
  \doi{\bibinfo{doi}{10.1007/978-3-540-45095-5\_8}}, \bibinfo{year}{2003}.

\bibitem[{Dikshit et~al.(2009)Dikshit, Zende, Bhatia, and
  Suri}]{Dikshit2009Convection}
\bibinfo{author}{B.~Dikshit}, \bibinfo{author}{G.~R. Zende},
  \bibinfo{author}{M.~S. Bhatia}, \bibinfo{author}{B.~M. Suri},
  \bibinfo{title}{Convection in molten pool created by a concentrated energy
  flux on a solid metal target}, \bibinfo{journal}{Physics of Fluids
  (1994-present)} \bibinfo{volume}{21}~(\bibinfo{number}{8})
  (\bibinfo{year}{2009}) \bibinfo{pages}{084105+}, ISSN
  \bibinfo{issn}{1070-6631}, \doi{\bibinfo{doi}{10.1063/1.3210763}}.

\bibitem[{Gilard and Brizzi(2005)}]{Gilard2005Slot}
\bibinfo{author}{V.~Gilard}, \bibinfo{author}{L.-E. Brizzi},
  \bibinfo{title}{{Slot Jet Impinging On A Concave Curved Wall}},
  \bibinfo{journal}{Journal of Fluids Engineering}
  \bibinfo{volume}{127}~(\bibinfo{number}{3}) (\bibinfo{year}{2005})
  \bibinfo{pages}{595+}, ISSN \bibinfo{issn}{00982202},
  \doi{\bibinfo{doi}{10.1115/1.1905643}}.

\bibitem[{Aidun and Martin(1997)}]{Aidun1997Effect}
\bibinfo{author}{D.~K. Aidun}, \bibinfo{author}{S.~A. Martin},
  \bibinfo{title}{Effect of sulfur and oxygen on weld penetration of
  high-purity austenitic stainless steels}, \bibinfo{journal}{Journal of
  Materials Engineering and Performance}
  \bibinfo{volume}{6}~(\bibinfo{number}{4}) (\bibinfo{year}{1997})
  \bibinfo{pages}{496--502}, ISSN \bibinfo{issn}{1059-9495},
  \doi{\bibinfo{doi}{10.1007/s11665-997-0121-1}}.

\bibitem[{Daub(2012)}]{Daub2012Erhohung}
\bibinfo{author}{R.~Daub}, \bibinfo{title}{Erh\"{o}hung der Nahttiefe beim
  Laserstrahl- W\"{a}rmeleitungsschwei{\ss}en von St\"{a}hlen}, Ph.D. thesis,
  \bibinfo{school}{Technische Universit\"{a}t M\"{u}nchen},
  \bibinfo{year}{2012}.

\bibitem[{Tanaka(2005)}]{Tanaka2005Effects}
\bibinfo{author}{M.~Tanaka}, \bibinfo{title}{Effects of surface active elements
  on weld pool formation using TIG arcs}, \bibinfo{journal}{Welding
  International} \bibinfo{volume}{19}~(\bibinfo{number}{11})
  (\bibinfo{year}{2005}) \bibinfo{pages}{870--876},
  \doi{\bibinfo{doi}{10.1533/wint.2005.3517}}.

\end{thebibliography}


\section*{References}

\begin{thebibliography}{1}

\end{thebibliography}

\end{document}